\pdfoutput=1
\documentclass[aps,prd,amsmath,floats,floatfix, twocolumn,
superscriptaddress,nofootinbib,showpacs]{revtex4-2}

\usepackage[T1]{fontenc}
\usepackage[utf8]{inputenc}
\usepackage{lmodern}
\usepackage[caption=false]{subfig}
\usepackage{verbatim}
\usepackage{multirow}
\usepackage{tikz}
\usetikzlibrary{decorations.pathmorphing}
\usetikzlibrary{decorations.markings}

\usepackage[dvipsnames, usenames]{xcolor}
\definecolor{linkcolor}{rgb}{0.0,0.3,0.5}
\usepackage[hypertexnames=false, unicode, colorlinks=true, linkcolor=linkcolor,
citecolor=linkcolor, filecolor=linkcolor,urlcolor=linkcolor,
pdfusetitle]{hyperref}

\usepackage[all]{hypcap}
\usepackage{graphicx}
\usepackage{xspace}
\usepackage{amssymb}
\usepackage[normalem]{ulem} 
\usepackage{bm} 
\usepackage{enumitem,amssymb}

\usepackage{microtype}
\usepackage{orcidlink}

\usepackage[english]{babel}
\usepackage{blindtext}

\graphicspath{%
  {figs/}%
}

\DeclareMathAlphabet{\mathpzc}{OT1}{pzc}{m}{it}

\newlist{todolist}{itemize}{2}
\setlist[todolist]{label=$\square$}

\begin{document}

\title{Emergent Turbulence in  Nonlinear Gravity}

\newcommand{\Cornell}{\affiliation{Cornell Center for Astrophysics
    and Planetary Science, Cornell University, Ithaca, New York 14853, USA}}
\newcommand\CornellPhys{\affiliation{Department of Physics, Cornell
    University, Ithaca, New York 14853, USA}}
\newcommand\Caltech{\affiliation{TAPIR 350-17, California Institute of
    Technology, 1200 E California Boulevard, Pasadena, CA 91125, USA}}
\newcommand{\Perimeter}{\affiliation{Perimeter Institute for Theoretical Physics, Waterloo, ON N2L2Y5, Canada}}
\newcommand{\AEI}{\affiliation{Max Planck Institute for Gravitational Physics
(Albert Einstein Institute), D-14476 Potsdam, Germany}}

\author{Sizheng Ma\orcidlink{0000-0002-4645-453X}}
\email{sma2@perimeterinstitute.ca}
\Perimeter

\author{Luis Lehner}
\email{llehner@perimeterinstitute.ca}
\Perimeter

\author{Huan Yang \orcidlink{0000-0002-9965-3030}}
 \email{hyangdoa@tsinghua.edu.cn} 
    \affiliation{Department of Astronomy, Tsinghua University, Beijing 100084, China}

\author{Lawrence E.~Kidder}
\Cornell

\author{Harald P. Pfeiffer}
\AEI

\author{Mark A. Scheel}
\Caltech
\hypersetup{pdfauthor={Ma et al.}}

\date{\today}

\begin{abstract}
Gravity in nonlinear and dynamical regimes underpins spectacular astrophysical phenomena and observable consequences, from the early universe to black hole collisions. In these extreme environments, {\em inverse energy cascades} --- mediated by nonlinear interactions --- may help explain the near scale-invariance of cosmic structure and the simplicity of gravitational waves from binary black hole mergers.
Yet the presence, characteristics, and generality of such interactions in full General Relativity remain largely unexplored. Here we show that two types of nonlinear interactions --- a four-mode and a three-mode interaction --- emerge in the fully nonlinear regime, and can indeed channel inverse energy cascades by inducing resonant and anti-damping (transient) nonlinear instabilities. 
We further demonstrate a ``laminar'' to ``turbulent'' transition for the largest-possible {angular} structure in General Relativity, whereas finer structures remain persistently turbulent. Our results reveal the impact and generality of these nonlinear interactions (instabilities), which can be key to understanding observations ranging from cosmological to kilometer scales.
We anticipate that our work will shed new light on nonlinear gravitational phenomena and their
consequences, such as constructing gravitational wave templates and testing General Relativity in the most extreme regime. Moreover, our work is a starting point for addressing nonlinear gravitational interactions using ideas and methods inspired by fluid dynamics. 
\end{abstract}

\maketitle

\section*{introduction}
Understanding truly nonlinear phenomena presents significant challenges. One prominent example is furnished by fluid dynamics, whose intrinsic complexity demands ingenious strategies to unravel its rich phenomenology.
A particularly powerful approach is to study systems continuously driven by an external source, thereby establishing a steady-state regime amenable to systematic analysis. Insights obtained in this regime often shed light on undriven or transient behavior.
One example is the identification of direct and inverse energy cascades in 3+1 and 2+1 dimensions, respectively~\cite{1941DoSSR..30..301K,10.1063/1.1762301,1980RPPh...43..547K},
which arise through mode couplings that both generate higher-order excitations and mediate energy flow across structures of different sizes, toward smaller (larger) scales with higher (lower) angular indices $\ell$ and frequencies.
This understanding has provided new avenues to both control and capture the complex dynamics exhibited by turbulent regimes,
e.g.~\cite{Castelvecchi2017,2023NatPh..19.1193M,2024Natur.627..515D,doi:10.1126/sciadv.ads5990}.

Gravity, as described by General Relativity (GR), remains only partially understood in the fully nonlinear regime. While major results have been obtained for stationary spacetimes, asymptotic structures, and perturbations around special solutions, the nonlinear and dynamical aspects still pose significant challenges at the level of first principles --- even though numerical simulations can now routinely model the collisions of compact binary systems and predict the resulting gravitational-wave (GW) signatures. This raises fundamental questions about the seemingly simple structure of these GWs, and how the system transitions from nonlinear to linear behavior during the post-merger stage of binary black hole (BBH) coalescence --- a linear regime crucial for BH spectroscopy \cite{Berti:2025hly}.


Our main interest here is to unravel this nonlinear regime, particularly in 3+1 dimensions. Hints of exciting phenomena have been obtained through perturbative analyses in convenient settings where linear perturbations persist long enough to channel interesting nonlinear behavior. Notably, \cite{Yang:2014tla,Iuliano:2024ogr} uncovered a parametric instability that preferentially transfers energy to larger scales.
These works, in turn, were motivated by the ``fluid-gravity duality'', in which certain gravitational systems in d+1 dimensions (within asymptotically anti-de Sitter spacetimes) are dual to (d-1)+1 relativistic hydrodynamics~\cite{Bhattacharyya:2007vjd}. In this framework, long-wavelength perturbations of BHs correspond, on the hydrodynamic side, to scenarios where turbulence can take place~\cite{VanRaamsdonk:2008fp,Carrasco:2012nf}.
Explicit illustrations of such behavior have been produced through numerical simulations
on both sides of the duality~\cite{Green:2013zba,Adams:2013vsa}. However, such duality requires the unphysical assumption of gravity with a negative cosmological constant.  

Taking inspiration from fluid dynamics,  here we explore nonlinear gravity in asymptotically flat spacetimes by 
introducing a {\em steady driving source} that gravitationally stirs the system through incoming GWs, and capturing the full nonlinear behavior through numerical simulations of Einstein's equations. 
Our approach establishes an analogous quasi–steady regime in gravity, providing a controlled setting to reveal nonlinear gravitational phenomena that would otherwise manifest transiently in non-driven scenarios. In doing so, it opens a new avenue for exploring nonlinear dynamics in GR --- one that complements traditional perturbative analyses by engaging nonlinear effects directly and borrowing conceptual guidance from fluid dynamics.

Our driving waves have prescribable amplitude, frequency, and spatial extent, allowing systematic exploration of the spacetime response. 
By varying these parameters, we quantitatively characterize energy transfer across frequencies and spatial scales, identify dominant interaction channels and their instabilities, and distinguish linear (laminar) from nonlinear (turbulent) behavior.
This framework provides a unified picture of how fully nonlinear GR drives inverse cascades and channels energy toward longer wavelengths, laying the groundwork for future analytical and phenomenological studies.



\section*{Driving Gravity}
\label{sec:driving}

We ``stir'' spacetimes by injecting (quasi-)steady GWs through the boundary of
a finite simulation domain. This is achieved by evolving Einstein’s equations in the Generalized Harmonic Formulation, discretized using pseudo-spectral methods with constraint-preserving boundary conditions. Further details are provided in the Supplementary Material (SM).


We study scenarios that initially correspond either to flat spacetime or 
a non-spinning, asymptotically flat BH\footnote{Linear perturbations in this regime decay faster
than in the spinning case, making it more difficult to excite nonlinear behavior.}, where the injected waves act as a sustained source of dynamical self-interaction and turbulent gravitational phenomena.
The driving is specified through a curvature scalar at the boundary \cite{Kidder:2004rw,Lindblom:2005qh,Ma:2023qjn,Ma:2024hzq}:
\begin{align}
    \Psi_0(t,\theta,\phi)=\sum_{\ell,m=0}A_{\ell m}(t) \times {}_{+2}Y_{\ell m}(\theta,\phi), 
\end{align}
where the angular part is described by spin-weighted spherical harmonics of spin weight $+2$. In the following, we inject a single mode $(\ell,m=0)$ at a time, considering $\ell=2$ or $6$ separately.
The former represents the longest angular structure allowed in GR; thus, there is no room to induce structure at lower $\ell$ values\footnote{$\ell=0$ and 1 correspond to mass and angular momentum changes.}, whereas for $\ell=6$, angular inverse cascades are possible.

The BH is located at the center of the domain with an initial
mass $M_i$. We adopt
\begin{align}
    A_{\ell,m=0}(t) = \begin{cases}
A_{i} \times t/t_0 \sin \omega t & \text{if } t \leq t_0,\\
A_{i} \sin \omega t  & \text{if } t > t_0,
\end{cases}
\label{eq:A_t_ramp_up_function}
\end{align}
where the injection amplitude is linearly ramped up to $A_{i}$ over $t_0= 600M_i$. This gradual turn-on is essential for stably simulating large $A_i$'s. 
The outer boundary is at $R_{\rm Bdry}=100M_{i}$. We have verified that its location does not affect conclusions. 

The frequency $\omega$ is fixed and real.
Wave absorption by the BH depends on the transmissivity of the potential barrier, which in linear theory rises sharply from near zero to unity as $\omega$ crosses the real part of the quasinormal-mode frequency (see SM). We find this trend persists even in our full nonlinear regime: low-$\omega$ waves are reflected, while for large $\omega$, accretion occurs, and the BH mass increases at a constant rate scaling as $A_{i}^2$. In the most nonlinear case, the BH mass increases by more than a factor of six. For intermediate frequencies near the quasinormal mode, the accretion rate gradually rises to a steady asymptotic value as the BH grows. Throughout, the BH spin remains consistent with zero, since the injected $(\ell=2,m=0)$ GWs carry no angular momentum.

The spacetime geometry is encoded in the metric. Its response to driving can be monitored via the ``tendicity'' scalar $\mathcal{E}$, which quantifies the tidal response experienced by orthogonal observers on a given worldtube  \cite{Owen:2010fa,Nichols:2011pu,Zimmerman:2011as,Zhang:2012jj,Nichols:2012jn}. This scalar is constructed from the Weyl tensor as:
\begin{equation}
    \begin{split}
        & \mathcal{E}=s^{\rho} s^{\sigma} C_{\rho\mu\sigma\nu}n^\mu n^\nu, 
    \end{split}
\end{equation}
where $n^a, s^a$ are unit normal vectors orthonormal to the spacetime foliation and the worldtube.
For initially flat spacetime, we adopt a worldtube at fixed coordinate radius; while in the BH case, we choose the apparent horizon. We examine how tendicity's harmonic components respond to the injected GW driver.

\begin{figure}[!htb]
    \centering
    \includegraphics[width=\linewidth]{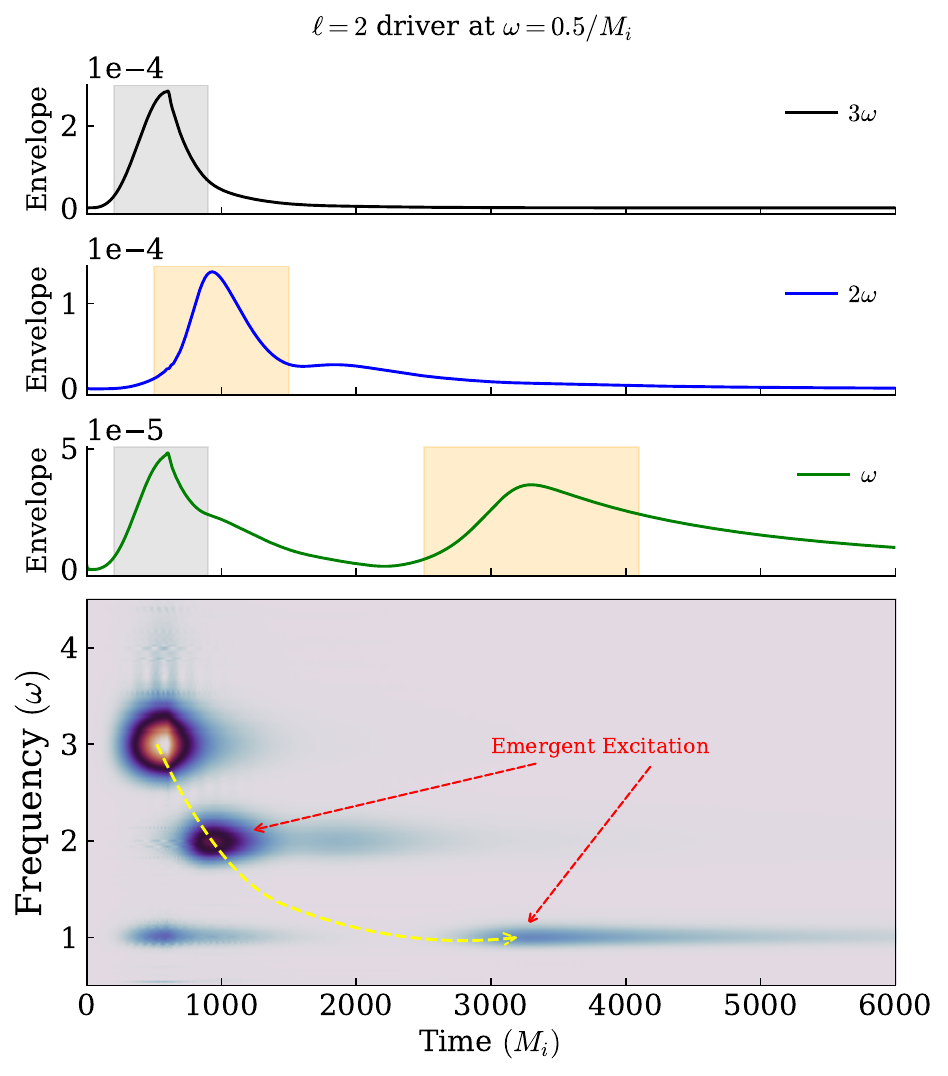}
    \caption{Stirring a non-spinning BH with a $\ell=2$ GW driver at fixed $\omega=0.5/M_i$ and $A_i=1.2\times10^{-3}$.
    The $\ell=6$ response of $\mathcal{E}$ exhibits three frequency components ($3\omega,2\omega,\omega$).
    Their amplitude envelopes are shown in the top three panels. The $3\omega$ and $\omega$ modes first arise through third-order couplings (gray), followed by nonlinear growth and saturation of $2\omega$ and $\omega$ (orange). Bottom: time–frequency representation exhibiting an inverse cascade: $3\omega\to 2\omega\to \omega$ (yellow dashed). }
    \label{fig:inverse_cascade_in_frequency_injecting_l2}
\end{figure}

\begin{figure}[!htb]
    \centering
      \includegraphics[width=\linewidth]{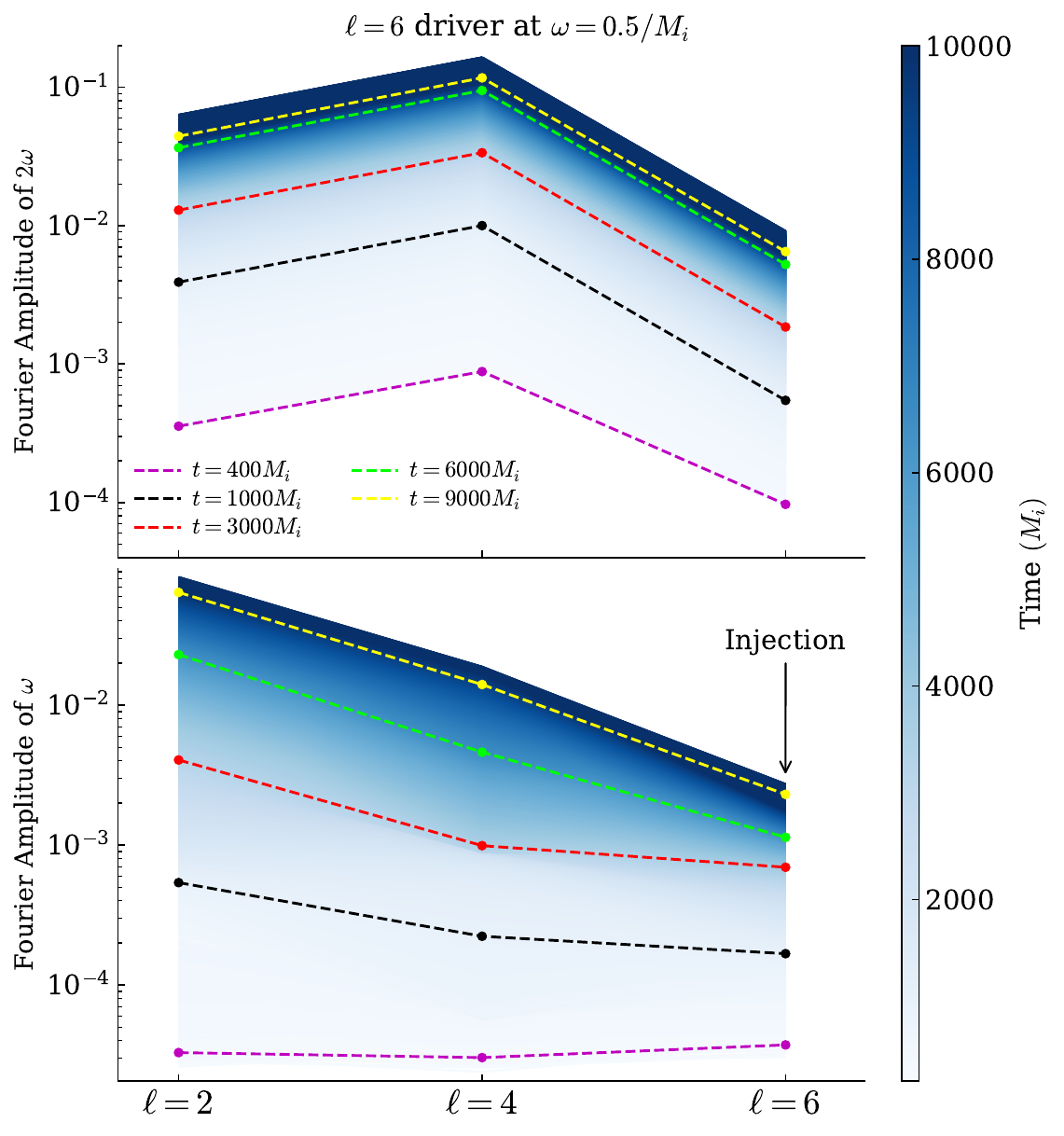}
    \caption{Stirring a non-spinning BH with a $\ell=6$ GW driver at $\omega=0.5/M_i$ and $A_i=4\times10^{-4}$.  Response across angular harmonics is measured continuously in time (in blue), with representative time slices shown (dot-dashed). Each harmonic contains $2\omega$ and $\omega$. The $2\omega$ component (top) arises from quadratic couplings and maintains a stable angular spectrum, whereas the $\omega$ mode (bottom) continuously redistributes its spectrum, yielding an inverse cascade toward lower $\ell$. }
    \label{fig:inverse_cascade_in_frequency_injecting_l6}
\end{figure}

\begin{figure}[!htb]
  \centering
  \includegraphics[width=\linewidth]{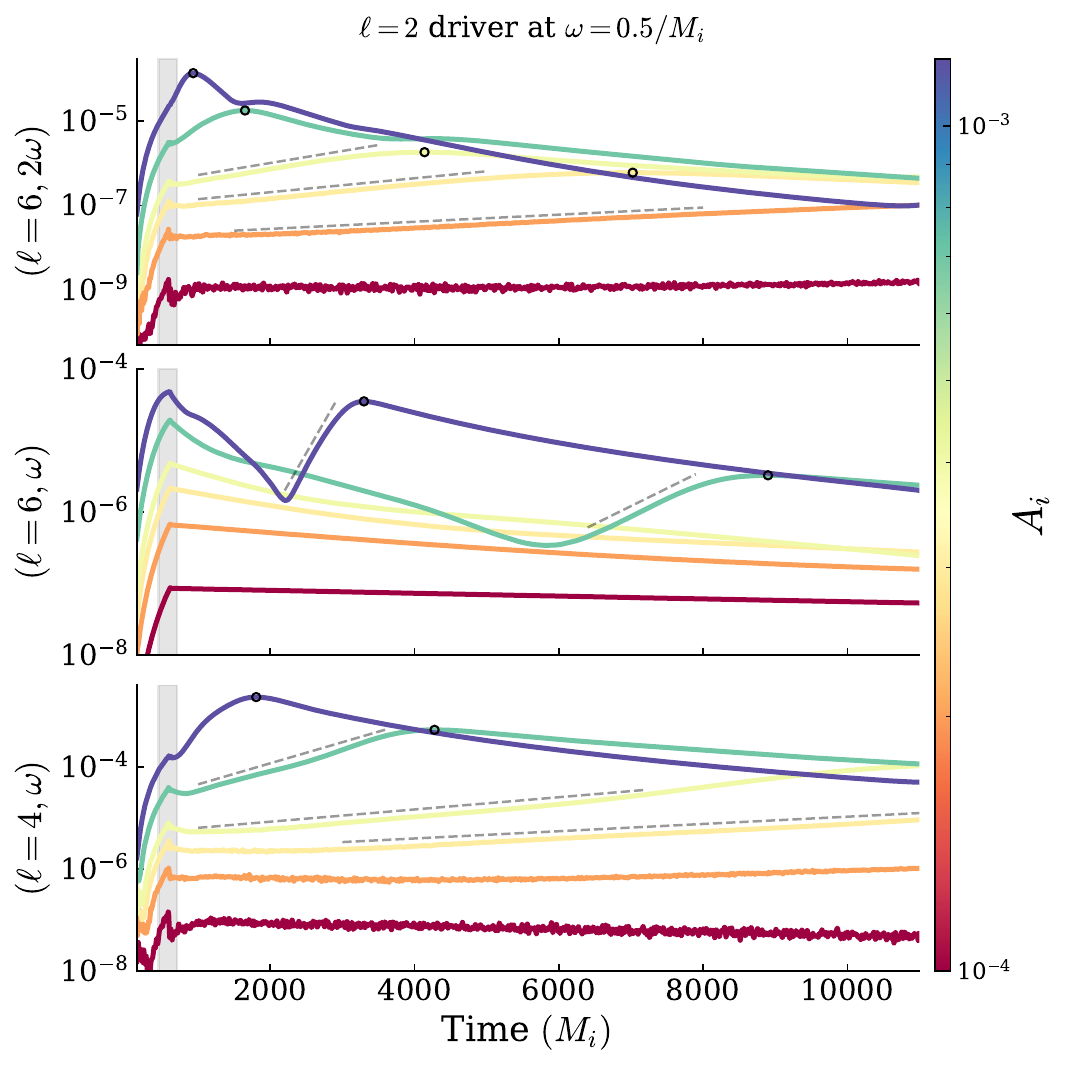} 
  \caption{Emergent instabilities in a stirred BH induced by $\ell=2$ GWs at $\omega=0.5/M_i$ with various amplitudes (colors).  Shown are amplitude envelopes of $\ell=6$ ($2\omega$ and $\omega$) and $\ell=4$ ($\omega$). Modes first undergo laminar excitations via fourth- and third-order couplings (gray). At low amplitudes, the structure persists; at higher amplitudes, new excitations emerge. For intermediate amplitudes, modes grow exponentially (gray dashed) before saturation (open circles). For $2\omega$ in $\ell=6$, the exponential growth rate scales as $A_i^2$ and saturation amplitude as $A_i^4$.}
  \label{fig:injecting_l2_various_amps}
\end{figure}

\begin{figure}[!htb]
  \centering
  \includegraphics[width=\linewidth]{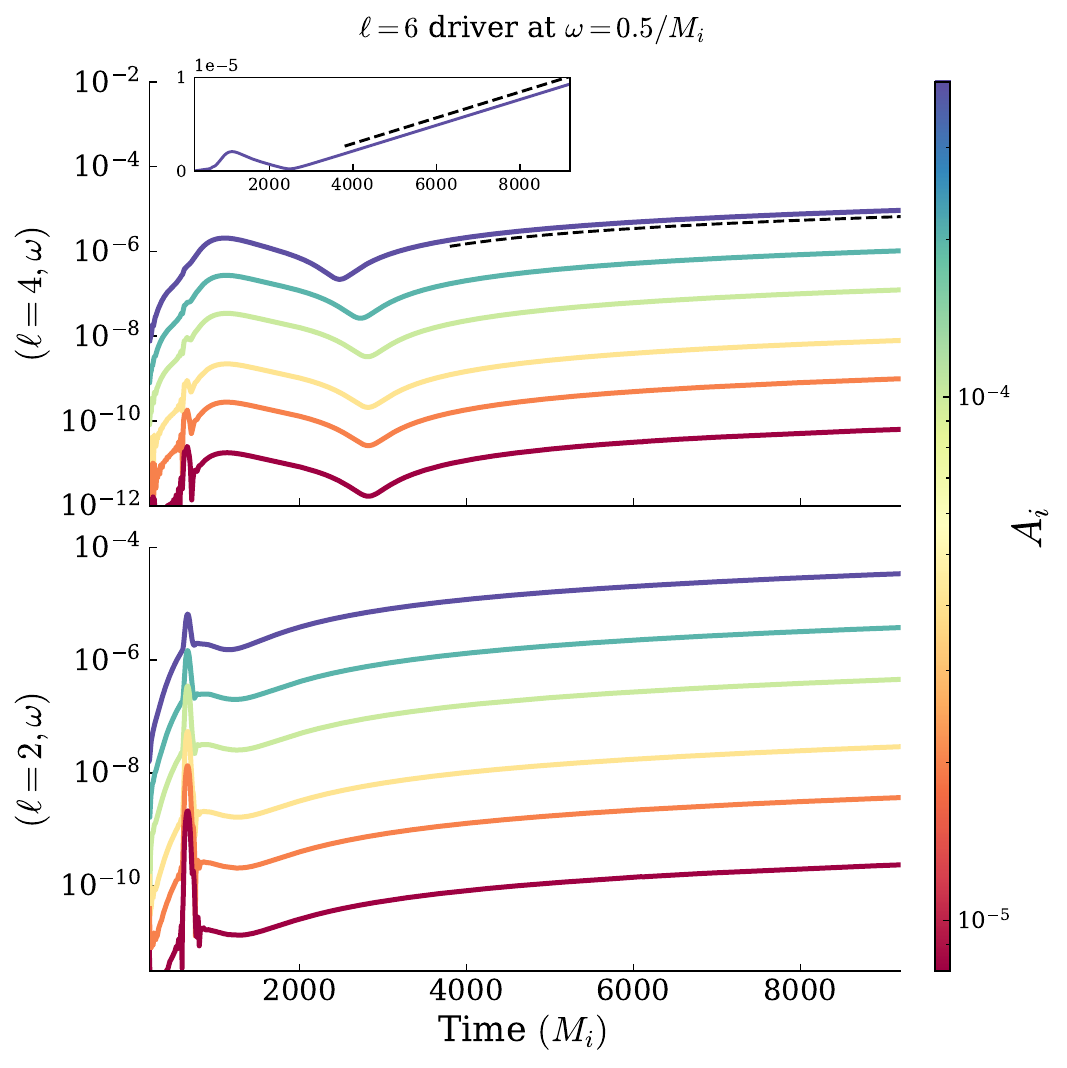}
  \caption{Emergent instabilities in a stirred BH induced by $\ell=6$ GWs at $\omega=0.5/M_i$ with various amplitudes (colors). The $\omega$ mode in $\ell=2$ and $4$ always grows linearly (dashed lines) regardless of $A_i$. The growth rate scales as $A_i^3$.}
  \label{fig:injecting_l6_various_amps}
\end{figure}

\section*{Results}
Our main conclusion is the emergence of temporal and spatial {\em inverse energy cascades} in nonlinear dynamical spacetimes. For a Schwarzschild BH, Fig.~\ref{fig:inverse_cascade_in_frequency_injecting_l2}
illustrates this effect while injecting an $\ell=2$ GW, with $\omega=0.5/M_i$ and $A_i=1.2\times10^{-3}$. The envelope of the response in the $\ell=6$ harmonic of $\mathcal{E}$ is shown, decomposed into its constituent frequency components. This harmonic first undergoes two nonlinear excitations at $\omega$ and $3\omega$ (gray region), arising from third-order couplings: $\omega=\omega+\omega-\omega$ and $3\omega=\omega+\omega+\omega$. Their peak amplitudes have been verified to scale as $A_i^3$. 
In the {\em weakly driven regime} these components evolve laminarily:  their response is \emph{instantaneous}, remains \emph{phase-locked} to the driving wave, and does not exchange energy with other modes. These features persist over time, and the same modes are excited regardless of the injection amplitude $A_i$.

Remarkably, for sufficiently strong injections, as in Fig.~\ref{fig:inverse_cascade_in_frequency_injecting_l2}, novel nonlinear excitations \emph{emerge} at $2\omega$ and $\omega$ sequentially in time (orange regions). 
These two excitations emerge, grow, and saturate after the laminar modes pass their peaks and the 
injected GW driver reaches its constant amplitude.
In particular, the $\omega$ component first undergoes a laminar excitation and then a secondary nonlinear emergence, resulting in a two-peaked structure in its temporal evolution.
The bottom panel of Fig.~\ref{fig:inverse_cascade_in_frequency_injecting_l2} further illustrates this phenomenology and provides more insights: the yellow dashed line, which connects the dominant laminar mode $3\omega$ to the subsequently emergent nonlinear modes $2\omega$ and $\omega$, indicates an inverse energy cascade towards smaller frequencies\footnote{We focus on $\omega M_i=0.5$. Similar nonlinear excitations also emerge at $\omega M_i=0.4$ and 0.6, but not at higher frequencies e.g., $\omega M_i=1.5$. Further details are provided in SM. }.

Figure \ref{fig:inverse_cascade_in_frequency_injecting_l6} considers an $\ell=6$ driver with $\omega=0.5/M_i$ and $A_i=4\times10^{-4}$, and shows the corresponding responses across various angular harmonics. Each angular harmonic contains two frequency constituents $\omega$ and $2\omega$. The $2\omega(=\omega+\omega)$ component arises via quadratic laminar couplings, whose amplitude scales as $A_i^2$. 
Its angular spectrum remains unchanged over time (top).
By contrast, the $\omega$ mode in $\ell=2$ and  $\ell=4$ emerges and grows. The angular spectrum evolves continuously (bottom), eventually developing greater power at larger angular scales (smaller $\ell$), indicating a spatial inverse cascade. Notably, this growth of $\omega$ occurs {\em regardless of $A_i$}, owing to the spatial structure of the $\ell > 2$ driver, which channels energy toward the largest GW angular scale, $\ell = 2$.

Figure~\ref{fig:injecting_l2_various_amps} provides a closer look at the $\ell=2$ injections, comparing across different driving strengths $A_i$. Laminar modes are excited at early times, including the fourth-order mode $2\omega(=\omega+\omega+\omega-\omega)$ in $\ell=6$ (top panel), as well as the third-order mode $\omega(=\omega+\omega-\omega)$ in $\ell=6,4$ (middle and bottom panels). These modes reach peak amplitudes just as the driver stabilizes (gray regions). The peak amplitudes are confirmed to scale as $A_i^4$ and $A_i^3$, respectively. For weak driving $(A_i \lesssim 2 \times 10^{-4})$, this early structure persists into later times, suggesting a ``laminar'' regime where modes are present but do not interact. As $A_i$ increases, a clear transition occurs: energy begins to cascade across modes, through secondary emergent excitations. For intermediate strength, the modes grow \emph{exponentially} (gray dashed lines) until saturation (open circles) and then decay. The growth becomes even faster for stronger amplitudes. Across various responses in $\ell=6$ at $2\omega$, the saturation amplitude scales as $A_i^4$, while the exponential growth rate scales as $A_i^2$, implying that multiple-order interactions are involved. 

Figure~\ref{fig:injecting_l6_various_amps} shows how energy inversely cascades into lower $\ell$ modes while injecting $\ell=6$ GWs (Fig.~\ref{fig:inverse_cascade_in_frequency_injecting_l6}). The $\omega$ component in $\ell=2$ and 4 increases \emph{linearly} over time, with the dashed line shown for reference. The inset plot further illustrates this linear growth. A systematic study finds the growth rate scales as $A_i^3$, implying third-order couplings. 

Notably, we find a similar linear growth in flat spacetime with $\ell=6$ drivers. However, its growth rate is three to four orders of magnitude smaller than around the BH. This suggests that the inverse cascade observed in Figs.~\ref{fig:inverse_cascade_in_frequency_injecting_l6} and \ref{fig:injecting_l6_various_amps} is not unique to BHs but instead reflects a more generic feature of nonlinear dynamics in GR, although the presence of BHs can significantly amplify the phenomenon.  In comparison, the emergent behavior shown in Figs.~\ref{fig:inverse_cascade_in_frequency_injecting_l2} and \ref{fig:injecting_l2_various_amps} does not manifest in the flat spacetime scenario.

Our results suggest that the temporal and spatial inverse cascades are driven by two types of nonlinear emergence. They occur after the excitation of laminar modes (gray regions in Figs.~\ref{fig:inverse_cascade_in_frequency_injecting_l2} and \ref{fig:injecting_l2_various_amps}) and when the external GW driver has stabilized to a constant amplitude. This sequence indicates that the novel excitations emerge from nonlinear interactions among these early excited laminar modes, reflecting underlying nonlinear instabilities. Below, we explore their possible physical origins. 

\begin{figure}[!htb]
  \centering
  \includegraphics[width=\linewidth]{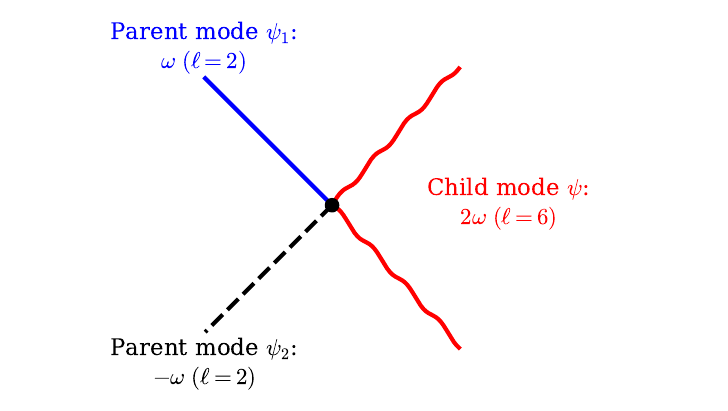}
  \includegraphics[width=\linewidth]{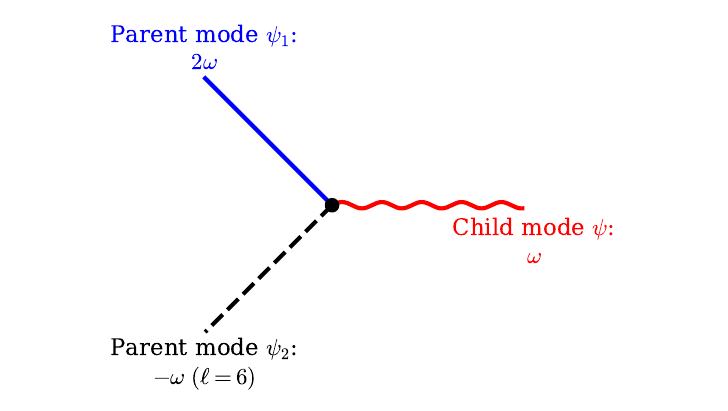}
  \caption{Feynman-type diagram illustrating the four-mode interaction (top) underlying the emergent excitations in Fig.~\ref{fig:injecting_l2_various_amps}; and the three-mode interaction (bottom) in Fig.~\ref{fig:injecting_l6_various_amps}. }
  \label{fig:detailedinteractions}
\end{figure}

In GR, the equations of motion schematically read $\mathcal{O}(\psi )= S(\psi)$, with $\mathcal{O}$ denoting a wave operator and $S$ a nonlinear source term. The instability unearthed in Figs.~\ref{fig:inverse_cascade_in_frequency_injecting_l2} and \ref{fig:injecting_l2_various_amps} originates from four-mode couplings, which can be expressed as
\begin{align}
    \mathcal{O}(\psi) =S^{(3)}(\psi_1, \psi_2, \psi_3),
\end{align}
where the trilinear source term $S^{(3)}$ depends on three parent modes (and complex conjugates). Focusing on the case that $\psi_3=\psi$ and $\omega_1+\omega_2=0$, the source may serve as an effective anti-damping term in the wave equation, yielding $\dot{A} \propto A (A_1 A_2)$, or equivalently $A \propto e^{{\rm const} (A_1A_2) t}$, where $A$ is the amplitude of the daughter mode $\psi$. For example, choosing the $\ell=2$ GW driver, we have $A_{1,2}=A_i$ and $\omega_{1,2}=\pm \omega$; the resulting daughter mode at $2\omega$ in $\ell=6$ then grows exponentially with a rate $\propto A^2_{i}$ before saturation. 
Figure \ref{fig:detailedinteractions} (top) shows a Feynman-type diagram for this four-mode coupling.

The second type of instability revealed in Figs.~\ref{fig:inverse_cascade_in_frequency_injecting_l6} and \ref{fig:injecting_l6_various_amps} can be explained by a three-mode coupling:
\begin{align}
    \mathcal{O}(\psi) =S^{(2)}(\psi_1, \psi_2),
\end{align}
where $S^{(2)}$ is a bilinear second-order source. Considering two parent modes $\psi_{1,2} \propto A_{1,2}e^{i \omega_{1,2}t}$, the daughter mode $\psi$ at $\omega_1+\omega_2$ is resonantly excited as $\psi \propto A_1A_2 t e^{i (\omega_1+\omega_2)t}$. With an $\ell=6$ driver, we have $ \omega_{1,2}=2\omega,-\omega$ and $A_{1,2}=A_i^2,A_i$, yielding $A \propto A^3_{i}$. This ``resonance''-type instability is analogous to a harmonic oscillator driven at its eigenfrequency. Figure \ref{fig:detailedinteractions} (bottom) illustrates the corresponding Feynman-type diagram.

\section*{Discussion}
We have unraveled nonlinear phenomena in GR in 3+1 settings. By stirring asymptotically flat spacetimes through quasi-steady GWs, we identified the crucial role of three- and four-mode interactions, arising from the full nonlinear structure of Einstein’s equations, in inducing resonant and anti-damping instabilities, which preferentially channel energy inversely toward larger scales. 
Our analysis demonstrates that the two instabilities are key drivers of emergent gravitational turbulence: the inverse energy cascade in both temporal and angular directions, together with the resonant and anti-damping instabilities revealed here, are essentially two sides of the same coin. Similar mechanisms, and their underlying
mathematical structure, are known in turbulent two-dimensional fluids in both driven and decaying turbulence (non-driven) scenarios~\cite{1980RPPh...43..547K}. While in fluid dynamics the inverse cascade is understood through enstrophy, its gravitational counterpart remains to be fully characterized.

Our findings could offer a new perspective on BBH dynamics: The pre-merger stage acts as a driven stage, while the merger initiates a non-driven/decaying phase that relaxes toward a laminar regime. Just as driven turbulence in fluids provides key insights into its decaying counterpart, our driven setup enables understanding how nonlinear gravitational dynamics settle into linear behavior. Especially, we find that the response to external driving favors the lowest angular structures and frequencies, consistent with theoretical and observational results for BBHs.  Future work could drive a BH into a quasi-steady state and then remove the forcing, thereby directly probing the gravitational analogue of the transition from driven to decaying turbulence. 

As discussed, our driven setup can establish an ``inertial-type'' regime and identify key nonlinear mechanisms that are expected to also operate in non-driven cases. This opens a new approach to studying nonlinear dynamics in GR. Meanwhile, the setup mimics astrophysical triple systems \cite{Cardoso:2021vjq,Santos:2025ass}, where a BH continuously receives GWs from a nearby
binary. Our observed turbulence suggests that similar phenomena could arise in such environments. Furthermore, the turbulence may connect to hidden integrability structures in GR~\cite{Jaramillo:2022oqn,Jaramillo:2024qjz}, nonlinear dynamics in the early universe~\cite{Galtier:2017mve,Galtier:2021ovg}, and nonlinear GW scattering in the post-Minkowskian framework, see e.g., \cite{Driesse:2024feo}.

Within the fluid-gravity correspondence, the unraveled turbulence extends beyond the classical duality, which needs a BH, a negative cosmological constant, and long-wavelength perturbations relative to the BH scale~\cite{Bhattacharyya:2007vjd}. 
Our results broaden and deepen earlier perturbative hints of such duality in specialized asymptotically flat settings~\cite{Yang:2014tla}.
This new understanding, which does not require a BH or cosmological constant,  offers a new platform for studying and comparing nonlinear mechanisms in gravity and fluid mechanics. It may also inspire explorations of emerging dualities such as celestial holography~\cite{Pasterski:2021raf} and the gravity-Carrollian hydrodynamics correspondence~\cite{Donnay:2019jiz}.

Although our steady gravitational driving and inverse cascades mark a concrete step toward the fluid–gravity correspondence, a fundamental distinction remains between them. Fluid nonlinearities arise from the principal (advective) terms, which allow characteristic crossings in the inviscid case; whereas in GR such crossings never occur, nonlinearity emerges from non-principal terms. In essence, gravity ``gravitates,'' focusing energy and mediating its nonlinear dynamics. Understanding how this distinction manifests within the correspondence is an intriguing direction.


\begin{acknowledgments}
We thank Luciano Combi for discussions during the initial stage of this work.
This work makes use of the Black Hole Perturbation Toolkit.
 LL acknowledges support from the Natural
Sciences and Engineering Research Council of Canada
through a Discovery Grant. LL also thanks financial support via the Carlo Fidani Rainer Weiss Chair at Perimeter
Institute and CIFAR. This research was supported in part
by Perimeter Institute for Theoretical Physics. Research
at Perimeter Institute is supported in part by the Government of Canada through the Department of Innovation,
Science and Economic Development and by the Province
of Ontario through the Ministry of Colleges and Universities. 
\end{acknowledgments}  

\def\bibsection{\section*{References}}
\bibliography{References}

\setcounter{page}{1}

\section*{Supplementary Material}
\section{frequency-dependent mass accretion}
\begin{figure}[!htb]
        \includegraphics[width=\linewidth]{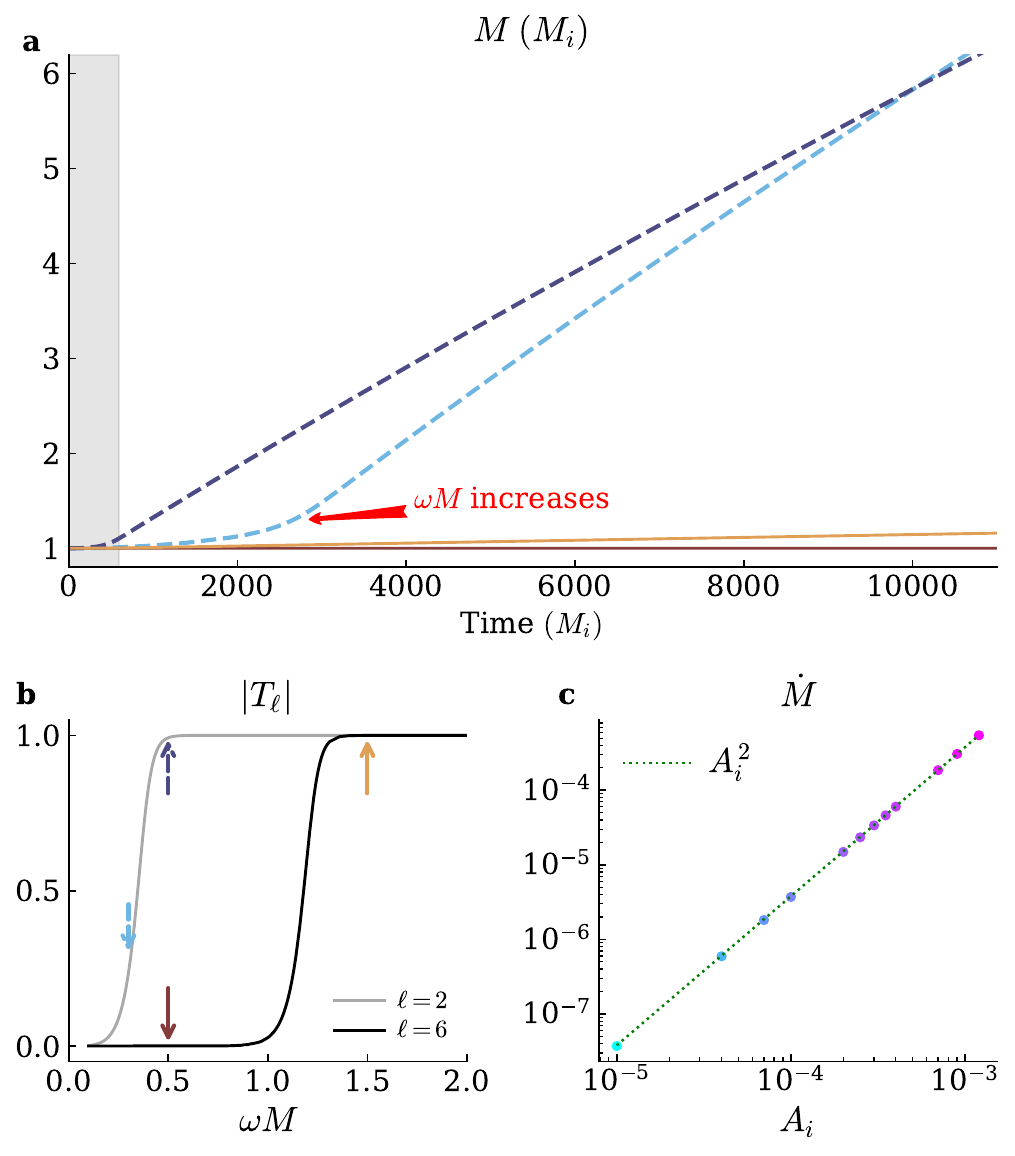}
     \caption{\textbf{Mass accretion induced by gravitational-wave drivers.} \textbf{a.} Time evolution of the black hole's Christodoulou mass $M$, normalized by its initial value $M_i$, for four representative cases: $\ell=2,\omega=0.5/M_i$ (dark-blue dashed), $\ell=2,\omega=0.3/M_i$ (light-blue dashed), $\ell=6,\omega=0.5/M_i$ (dark-brown solid), and $\ell=6,\omega=1.5/M_i$ (light-brown solid). \textbf{b.} Transmissivity of a Schwarzschild black hole for $\ell=2$ (gray) and $\ell=6$ (black), computed using Black Hole Perturbation Toolkit \cite{BHPToolkit}. Arrows indicate the four drivers used in \textbf{a}. \textbf{c.} Accretion rate induced by the $\ell=2,\omega=0.5/M_i$ driver as a function of injection amplitude $A_i$, which shows a clear quadratic dependence. }
    \label{fig:new_mass_change_suppl}
\end{figure}

As shown in Fig.~\ref{fig:new_mass_change_suppl}\textbf{\textcolor{linkcolor}{a}}, we investigate the time evolution of the black hole's Christodoulou mass, $M$, under several representative scenarios. In each case, the driving gravitational wave has a fixed frequency, expressed in units of the black hole’s initial mass $M_i$.
The driver's amplitude is gradually ramped up at early times (gray region) and then held constant thereafter.

Three distinct mass accretion behaviors are observed:
\begin{enumerate}
    \item For $(\ell=2,\omega=0.5/M_i)$ and $(\ell=6,\omega=1.5/M_i)$ --- shown as the dark-blue dashed and light-brown solid curves --- the black hole mass grows linearly with time, indicating a steady accretion regime --- albeit with different rates.
    \item For $(\ell=6,\omega=0.5/M_i)$ --- dark-brown solid curve --- the mass remains constant, signaling no accretion.
    \item For $(\ell=2,\omega=0.3/M_i)$ --- light-blue dashed curve --- the growth rate is initially low and time-dependent, but the system eventually transitions into a steady accretion phase at late times.
\end{enumerate}



The distinct accretion behaviors observed in our nonlinear simulations can be largely understood via black hole linear perturbation theory. Specifically, considering the Zerilli equation \cite{Zerilli:1970se}
\begin{align}
    \left(\frac{\partial^2}{\partial r_*^2}-\frac{\partial^2}{\partial t^2}-V_{\ell}\right)Z_{\ell}=0,
\end{align}
our setup is consistent with its \emph{in}-mode solution, whose boundary conditions are given by:
\begin{equation}
Z_{\ell} = \begin{cases}
e^{-i\omega (t+r_*)}+R_{\ell}e^{-i\omega (t-r_*)}, & {\rm as} \quad r_*\to +\infty, \\
T_{\ell} e^{-i\omega (t+r_*)}, & {\rm as} \quad r_*\to -\infty. 
\end{cases}
\label{eq:in_solution}
\end{equation}
Here, a gravitational wave is emitted toward the Schwarzschild black hole from past null infinity. Part of the wave is reflected, while the transmitted portion tunnels through the curvature potential and is absorbed by the horizon. As shown in Fig.~\ref{fig:new_mass_change_suppl}\textbf{\textcolor{linkcolor}{b}}, the transmission coefficient $|T_\ell|$ is nearly zero at low frequencies and increases toward unity as $\omega M$ increases. The transition happens around the real part of the corresponding fundamental quasinormal mode frequency, i.e., 0.374 for $\ell=2$ and 1.21 for $\ell=6$. 

\begin{figure*}[!htb]
        \includegraphics[width=0.95\linewidth]{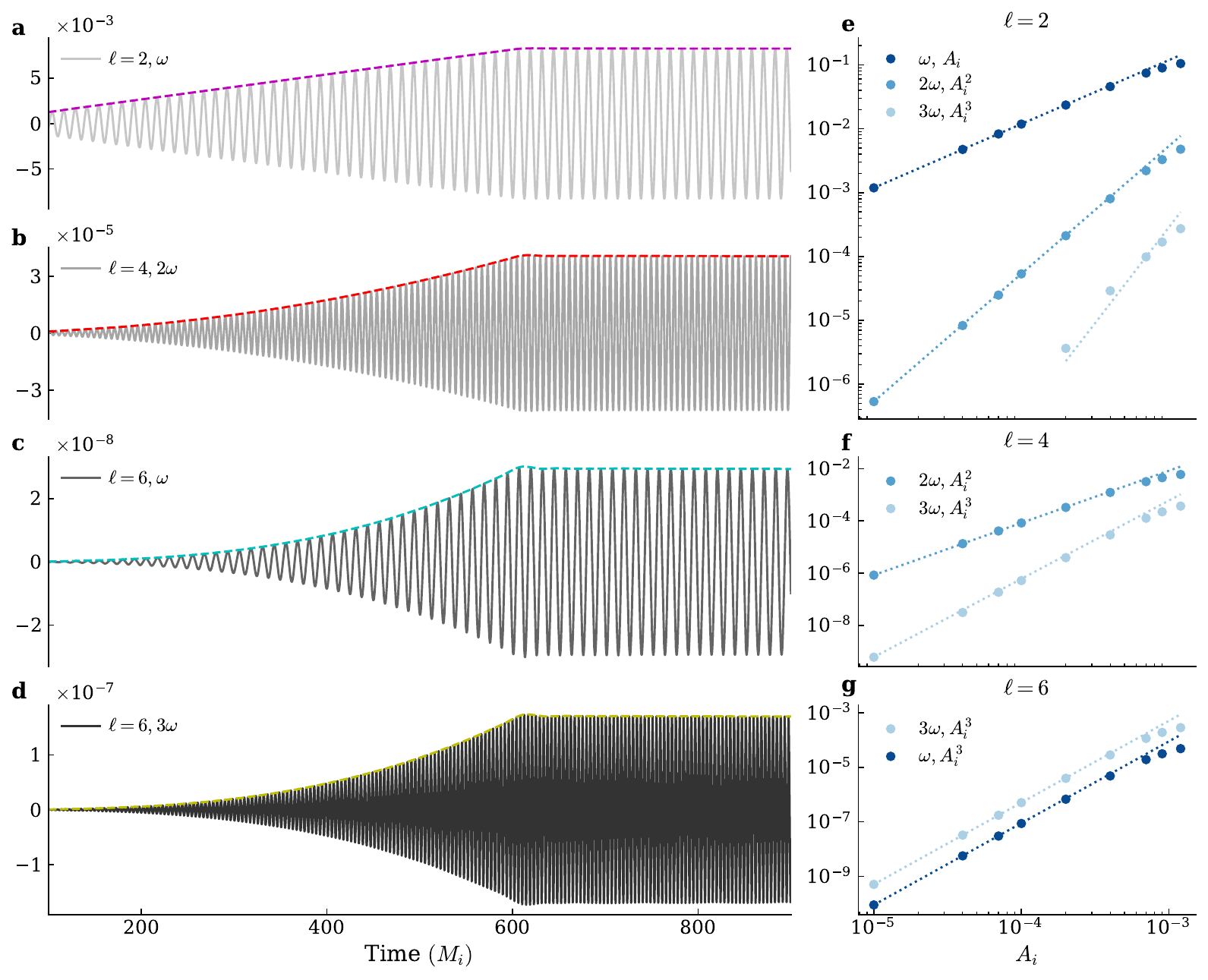}
    \caption{\textbf{Excitation of laminar modes by $\ell=2$ drivers}. \textbf{a-d.} Time evolution of the dominant frequency components in $\ell=2,4$, and 6 harmonics of the tendicity $\mathcal{E}$, induced by a driver with $\omega=0.5/M_i$ and $A_i=7\times10^{-5}$. Colored dashed curves highlight the corresponding envelopes. \textbf{e-g.} Measured mode amplitudes as functions of the injection strength $A_i$. The linear-, second-, and third-order modes scales as $A_i$, $A_i^2$, and $A_i^3$, as expected.  }
    \label{fig:injecting_l2_weak_injection_td_response_zoom_in}
\end{figure*}

In scenario (1), both waves have relatively high frequencies, as indicated by the dark-blue dashed arrow and the light-brown solid arrow in Fig.~\ref{fig:new_mass_change_suppl}\textbf{\textcolor{linkcolor}{b}}. Their transmission coefficients are close to unity, allowing nearly all of the wave energy to be absorbed. The black hole thus accretes efficiently and steadily, with its mass increasing linearly in time. In scenario (2), shown by the dark-brown solid arrow in Fig.~\ref{fig:new_mass_change_suppl}\textbf{\textcolor{linkcolor}{b}}, the frequency $\omega=0.5/M_i$ is much lower than the quasinormal mode frequency for $\ell=6$. The transmission coefficient is near zero, and the wave is almost entirely reflected --- leading to negligible mass growth. Scenario (3) corresponds to the light-blue dashed arrow in Fig.~\ref{fig:new_mass_change_suppl}\textbf{\textcolor{linkcolor}{b}}, where the transmissivity is initially small $(\lesssim0.3)$ but non-negligible. Only a fraction of the wave is absorbed at early times, resulting in a slow, time-dependent accretion rate. However, as the black hole mass $M$ increases, so does the product $\omega M$, the system moves to the right along the transmission curve\footnote{We emphasize that $M$ here refers to the instantaneous Christodoulou mass of the black hole, which evolves during the simulation. This should be distinguished from the fixed initial mass $M_i$. The injected frequency $\omega M_i$ remains constant throughout.}. Eventually, $|T_\ell|$ asymptotes to unity, and nearly all the wave energy is absorbed. At this stage, the system transitions to a steady accretion regime with a constant mass growth rate.

We also confirm that the steady accretion rate scales as $A_i^2$, where $A_i$ is the amplitude of the injected wave. This feature is consistent with the quadratic dependence of wave energy on amplitude. Figure \ref{fig:new_mass_change_suppl}\textbf{\textcolor{linkcolor}{c}} illustrates this scaling for the case $\ell=2$ and $\omega=0.5/M_i$.

In addition to mass growth, we find that the black hole's spin remains approximately zero $(\sim10^{-13})$ throughout simulations, as expected for injections of $(\ell=2,m=0)$ waves.
Taken together, these results provide a strong internal consistency check for our numerical setup. They demonstrate that the black hole absorbs gravitational waves and exhibits frequency-dependent accretion behavior in excellent agreement with expectations from linear perturbation theory. When the transmission coefficient approaches unity, the black hole enters a steady accretion regime.

\section{Case study: $\ell=2$ driver }
\label{sec:Nonlinear_parametric_instability}
As discussed in the main text, injecting strong $(\ell=2,m=0)$ gravitational waves leads to the emergence of a four-mode coupling. This coupling drives a laminar-to-turbulent transition as the injection amplitude $A_i$ increases. In the laminar regime, nonlinear modes are excited and persist with negligible interaction, while in the turbulent regime, certain modes undergo secondary excitation (anti-damping instability), yielding an inverse energy cascade.

In the following, we elaborate on these phenomena by presenting additional results and interpretations for both the weak regime (Sec.~\ref{subsec:parametric_instability_weak_regime}) and the strong regime (Sec.~\ref{subsec:l2_strong_regime}). In particular, the discussion of the weak regime helps build intuition about our setup and draw on insights from black hole linear perturbation theory and tidal phenomena.

\subsection{Weak regime}
\label{subsec:parametric_instability_weak_regime}
The main results are summarized in Fig.~\ref{fig:injecting_l2_weak_injection_td_response_zoom_in}, where the gravitational-wave driver has amplitude $A_{i}=7\times10^{-5}$ and  frequency $\omega=0.5/M_{i}$. This frequency lies above the fundamental quasinormal mode, allowing the injected wave to penetrate the curvature barrier and reach the horizon. The tendicity $\mathcal{E}$ has nontrivial response in the $(\ell={\rm even}, m=0)$ harmonics, with each harmonic composed of frequency components at integer multiples of the injection frequency $\omega$. Figure \ref{fig:injecting_l2_weak_injection_td_response_zoom_in}\textbf{\textcolor{linkcolor}{a-d}} show that each frequency mode closely tracks the behavior of the injected wave: its amplitude gradually increases prior to $t_0=600M_i$, and then settles into a steady oscillation once the injection stabilizes, with no further nonlinear evolution. This prompt response is reminiscent of the adiabatic tidal response in neutron stars subjected to external fields.

The presence and behavior of these modes can be understood through standard perturbative couplings. At $n$-th order in perturbation theory, a mode arises from self-coupling of the linear perturbation --- characterized by frequency $\omega$ and angular index $\ell$ --- through $n$ successive interactions. Formally, one can write 
\begin{align}
\underbrace{(\omega,\ell)\otimes\ldots(\omega,\ell)}_{n}.
\end{align}
In our case, they lead to:
\begin{itemize}
    \item Linear order: $\omega$ in $\ell=2$, with amplitude $\propto A_{i}$.
    \item Second order: $2\omega(=\omega+\omega)$ in $\ell=2,4$, with amplitude $\propto A_{i}^2$.
    \item Third order: $3\omega(=\omega+\omega+\omega)$ and $\omega(=\omega+\omega-\omega)$ in $\ell=2,4,6$, with amplitudes $\propto A_{i}^3$.
\end{itemize}
These amplitude scaling relations are verified in Fig.~\ref{fig:injecting_l2_weak_injection_td_response_zoom_in}\textbf{\textcolor{linkcolor}{e-g}}, where the measured mode amplitudes match the expected power laws across two orders of magnitude in $A_{i}$. Slight deviations appear for large injections $(A_{i} \gtrsim 7\times10^{-4})$, when the secondary excitation (anti-damping instability) emerges.

The nonlinear daughter modes are direct analogs of the quadratic quasinormal modes recently studied in the literature \cite{Berti:2025hly}. Their properties have been investigated using black hole second-order perturbation theory, see e.g. \cite{Ma:2024qcv,Khera:2025cls}. In particular, \cite{Ma:2024qcv} showed a simple angular selection rule governing quadratic couplings $\ell_1\otimes\ell_2\to\ell_3$ in Schwarzschild black holes [see Eqs.~(77) and (78) therein]. In the present context, the coupling coefficient is determined by the $3-j$ symbol $\begin{pmatrix}
    \ell_1 & \ell_2 & \ell_3 \\
    0 & 0 & 0
\end{pmatrix}$, which is nonzero when
\begin{align}
|\ell_1 - \ell_2| \leq \ell_3 \leq \ell_1 + \ell_2. \label{eq:angular_selection_rule}
\end{align}
At third order, the angular selection rules can be formulated in terms of Racah W-coefficients \cite{Redondo-Yuste:2025hlv}.
This structure naturally suggests that a $n$th-order nonlinear interaction driven by a $\ell=2$ mode generically populates multiple harmonics up to $\ell_{\max}=2n$.
A direct consequence is that the $2\omega$ component in $\ell=6$ (see discussions below in Sec.~\ref{subsec:injecting_l2_various_frequencies}) cannot arise at quadratic order $(2\omega=\omega+\omega)$, but must instead originate from a fourth-order process: $2\omega=\omega+\omega+\omega-\omega$.



\begin{figure*}[!htb]
        \includegraphics[width=0.49\linewidth]{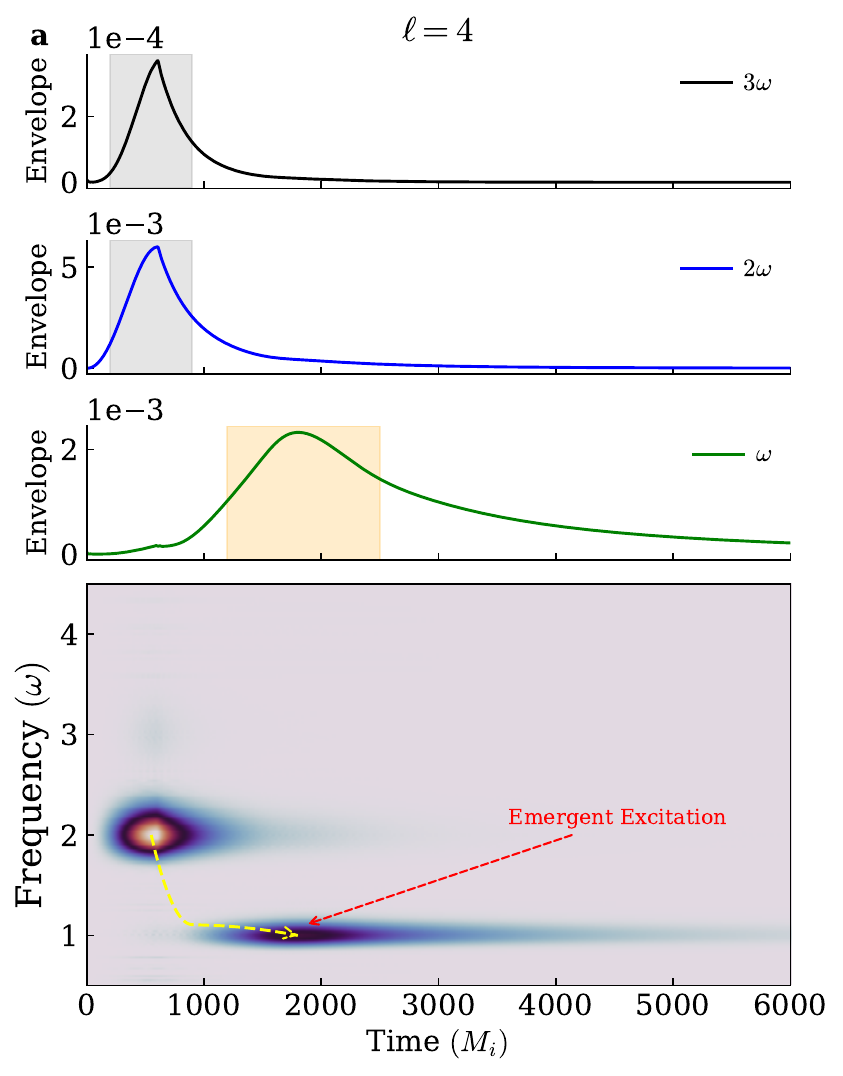}
        \includegraphics[width=0.49\linewidth]{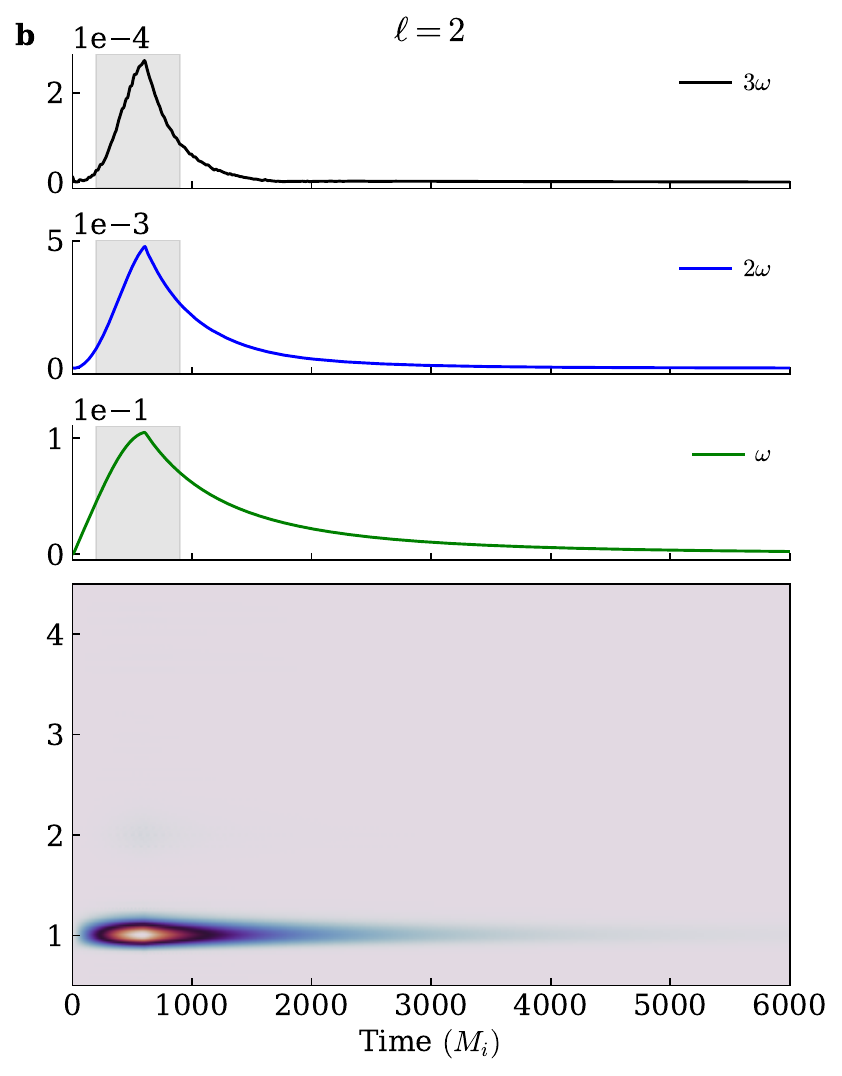}
    \caption{\textbf{Black hole's responses to a strong $\ell=2$ driver}. The driver has frequency $\omega=0.5/M_i$ and amplitude $A_i=1.2\times10^{-3}$. Responses are decomposed into their constituent frequency components, whose amplitude envelopes are shown. \textbf{a.} Response in $\ell=4$ of $\mathcal{E}$, featuring two laminar modes $3\omega,2\omega$ and an emergent mode $\omega$. The laminar modes peak when the gravitational-wave driver stabilizes (gray-shaded regions), whereas $\omega$ emerges and saturates later (orange-shaded region). The yellow dashed line in the bottom time–frequency plot highlights an inverse cascade toward lower frequency. \textbf{b.}  Response in $\ell=2$ of $\mathcal{E}$, including three laminar modes $3\omega$, $2\omega$, and $\omega$. No further nonlinear excitations are observed.}
\label{fig:inverse_cascade_in_frequency_injecting_l2_response_l4}
\end{figure*}

\subsection{Strong regime}\label{subsec:l2_strong_regime}
For our investigation, we adopt the same strong gravitational-wave driver as in the main text, with  $\omega=0.5/M_i$ and $A_i=1.2\times10^{-3}$. The corresponding mass evolution is shown as the dark-blue dashed curve in Fig.~\ref{fig:new_mass_change_suppl}\textbf{\textcolor{linkcolor}{a}}, where the black hole becomes six times bigger at the end of the simulation.

Figure \ref{fig:inverse_cascade_in_frequency_injecting_l2_response_l4} complements our discussions in the main text. Similar to the $\ell=6$ response, where $2\omega$ and $\omega$ emerge sequentially at late times, Fig.~\ref{fig:inverse_cascade_in_frequency_injecting_l2_response_l4}\textbf{\textcolor{linkcolor}{a}} shows that in $\ell=4$, the $\omega$ component is nonlinearly excited (orange region) after the laminar modes $2\omega$ and $3\omega$ (gray regions) have peaked and the driver has stabilized. As discussed in Sec.~\ref{subsec:parametric_instability_weak_regime} and Fig.~\ref{fig:injecting_l2_weak_injection_td_response_zoom_in}, these two laminar modes arise from second- and third-order couplings, respectively. The yellow dashed curve in the bottom panel of Fig.~\ref{fig:inverse_cascade_in_frequency_injecting_l2_response_l4}\textbf{\textcolor{linkcolor}{a}} connects the dominant laminar mode $2\omega$ with the later-emerging mode $\omega$, highlighting an inverse cascade.

The time evolution of $\omega$ in $\ell=4$ for a variety of injection amplitudes is presented in the main text. There, we showed that the behavior of this mode closely parallels the later-emerging $\omega$ and $2\omega$ in $\ell=6$: it does not grow until $A_{i}$ exceeds a threshold. For intermediate driver strengths, the mode grows exponentially; whereas for stronger injections, the growth deviates from exponential and saturates more rapidly.

Figure \ref{fig:inverse_cascade_in_frequency_injecting_l2_response_l4}\textbf{\textcolor{linkcolor}{b}} displays the response in $\ell=2$, which consists of three laminar modes $\omega$ (first-order), $2\omega$ (second-order), and $3\omega$ (third-order). Each mode peaks when the driver settles to a constant amplitude (gray regions), and does not exhibit any secondary excitations thereafter. This behavior supports our conclusion that \emph{gravity favors inverse cascades}: the dominant frequency component $\omega$ in $\ell=2$ already occupies the lowest available frequency, leaving no lower-frequency room for further energy transfer.

\begin{figure}[!htb]
        \includegraphics[width=\linewidth]{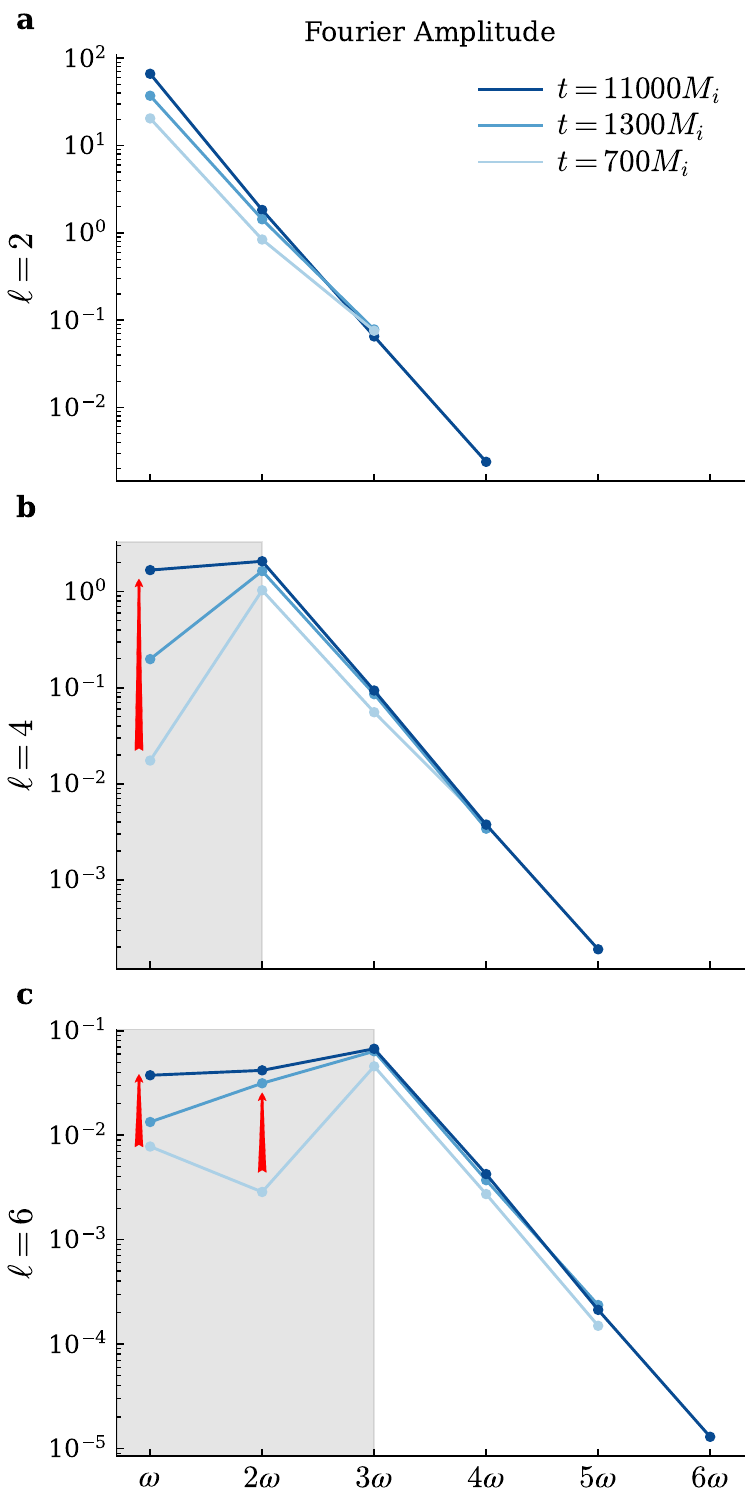}
    \caption{\textbf{Fourier spectra of the black hole's responses to a strong $\ell=2$ driver.}  The driver has frequency $\omega=0.5/M_i$ and amplitude $A_i=1.2\times10^{-3}$. Responses are shown for $\ell=2$ (\textbf{a}), $\ell=4$ (\textbf{b}), and $\ell=6$ (\textbf{c}), each over three successive time intervals. Red arrows mark the direction of spectral flow; grey shading indicates regions of inverse cascade.}
    \label{fig:FFT_injecting_l2}
\end{figure}

Our discussions in the main text and here phenomenologically suggest that the early-excited laminar modes serve as energy reservoirs to trigger secondary excitations (instability). One piece of supporting evidence is that these laminar modes begin to decay after reaching their peak amplitudes, whereas they remain constant in the weak-driving regime (see Fig.~\ref{fig:injecting_l2_weak_injection_td_response_zoom_in}). This contrast implies a transfer of energy from the laminar modes into later-emerging ones. To complement the earlier time-frequency analyses, Fig.~\ref{fig:FFT_injecting_l2} presents the Fourier amplitudes of frequency components in $\ell=2,4$ and $6$. We compute the spectra over three time intervals to track the progressive development of nonlinear excitations:
\begin{itemize}
    \item $t\in [0, 700M_{i}]$: The laminar modes are driven to their peak amplitudes.
    \item $t\in [0, 1300M_{i}]$: The secondary instability channels are active but have not yet saturated.
    \item $t\in [0, 11000M_{i}]$: All nonlinear instability channels have well saturated.
\end{itemize}
Within the time interval $[0, 700M_{i}]$, the amplitude distributions are qualitatively consistent with expectations from the normal perturbation perspective. As discussed in Sec.~\ref{subsec:parametric_instability_weak_regime}, in $\ell=2$, each frequency multiple $n \omega$ ($n\in {\rm integer}$) arises from a $n$-th order coupling, whose amplitude decreases exponentially with $n$. In $\ell=4$ (Fig.~\ref{fig:FFT_injecting_l2}\textbf{\textcolor{linkcolor}{b}}), the dominant frequency component is $2\omega$, originating at second order. The $\omega$ and $3\omega$ components both appear at third order, and thus have comparable amplitudes, together forming a ``\rotatebox[origin=c]{180}{v}'' shape.
Higher multiples arise from progressively higher-order interactions and show correspondingly smaller amplitudes.
Finally, in $\ell=6$ (Fig.~\ref{fig:FFT_injecting_l2}\textbf{\textcolor{linkcolor}{c}}), the leading third-order modes are $\omega$ and $3\omega$, followed by two fourth-order modes $2\omega$ and $4\omega$. Modes arising from the same perturbative order have comparable amplitudes, producing a ``\reflectbox{N}'' shape in the spectrum.

Moving on to the interval $t\in [0, 1300M_{i}]$, the spectrum of $\ell=2$ remains qualitatively similar to its earlier state. We note that the Fourier amplitudes of $\omega$ and $2\omega$ appear slightly stronger even though their instantaneous amplitudes decrease (see Fig.~\ref{fig:inverse_cascade_in_frequency_injecting_l2_response_l4}\textbf{\textcolor{linkcolor}{b}}). This is because the Fourier transformation integrates over the entire signal history --- extending the window in time naturally accumulates more signal content. In $\ell=4$, the $\omega$ component undergoes the secondary excitation as described in Fig.~\ref{fig:inverse_cascade_in_frequency_injecting_l2_response_l4}\textbf{\textcolor{linkcolor}{a}}, resulting in an upward shift of its Fourier amplitude. By contrast, the $2\omega$ and higher multiples remain dominated by the initial laminar excitation and show little change. In $\ell=6$, the $2\omega$ mode also experiences the nonlinear growth, and the combined spectrum of $\omega,2\omega,3\omega$ roughly aligns along a straight line (using a logarithmic scale). The $3\omega$ and higher multiples, on the contrary, remain close to their original laminar levels.

Finally, for  $t\in [0, 11000M_{i}]$, the Fourier amplitude of $\omega$ in $\ell=4$ becomes comparable to that of $2\omega$, forming a clear plateau at the low-frequency end of its spectrum. Similarly, in $\ell=6$, the spectrum develops a plateau extending up to $3\omega$ after the $\omega$ mode undergoes its secondary excitation. By contrast, the high-frequency portion of the spectra remains largely unchanged, retaining the exponential decay of the initial laminar excitation. In the figure, we shade the inverse-cascade region in gray and use red arrows to indicate the spectral flow. 

\begin{figure}[!htb]
        \includegraphics[width=\linewidth]{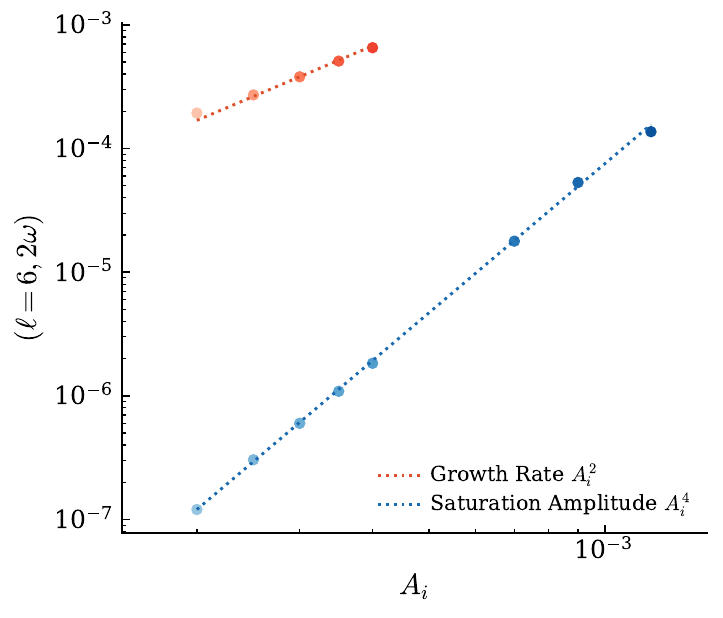}
    \caption{Measured growth rate (red) and saturation amplitude (blue) of $2\omega$ in $\ell=6$, shown as functions of the injection amplitude for the $\ell=2$ gravitational-wave driver. Dotted lines indicate the $A_i^2$ and $A_i^4$ scaling for comparison.}
    \label{fig:injecting_l2_parametric_response_l4_l6_vs_amp}
\end{figure}

To close this subsection, Figure \ref{fig:injecting_l2_parametric_response_l4_l6_vs_amp} presents the measured exponential growth rate (red) and saturation amplitude (blue) of $2\omega$ in $\ell=6$, as functions of $A_i$. The growth rate can be reliably extracted only within an intermediate range of $A_{i}$: at large amplitudes, the excitation deviates from exponential growth, while at smaller amplitudes, no growth occurs. Within this range, the growth rate scales as $A_{i}^2$. We thus attributed this phenomenon to the four-mode coupling introduced in the main text. The saturation amplitude follows $A_{i}^4$, indicating that it is governed by a fourth-order process. A detailed investigation of this saturation mechanism is left for future work.

\subsection{Injection at various frequencies}
\label{subsec:injecting_l2_various_frequencies}
Our investigation in Sec.~\ref{subsec:l2_strong_regime} focused on a specific injection frequency $\omega M_{i}=0.5$, chosen somewhat arbitrarily but intentionally close to and slightly above the $\ell=2$ fundamental quasinormal mode frequency.
One motivation for this choice was to ensure that a large portion of the injected wave reaches the horizon. In what follows, we demonstrate that the key features of the emergent instability persist for other low-frequency injections.

\begin{figure}[!htb]
        \includegraphics[width=\linewidth]{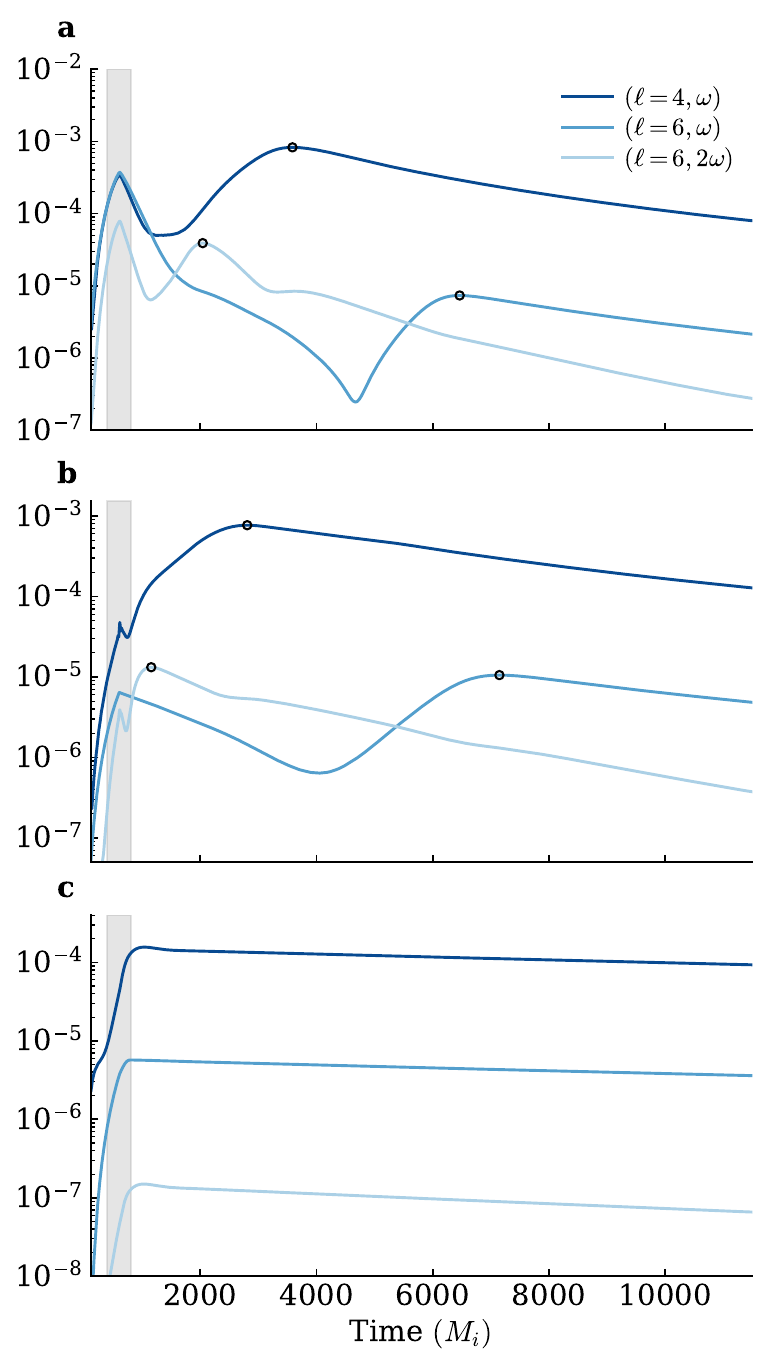}
    \caption{\textbf{Black hole's responses to strong $\ell=2$ drivers at three frequencies.} Injection frequencies are $\omega M_i=$ 0.4 (\textbf{a}), 0.6 (\textbf{b}), and 1.5 (\textbf{c}). Only modes expected to undergo emergent instabilities are considered. For each case, the amplitude envelope of the response is shown. Gray-shaded regions indicate the period during which the injected gravitational-wave driver stabilizes to a constant amplitude. Instabilities are observed in the first two cases, with saturation marked by open circles. In the third case, all modes remain in the laminar regime and no instabilities are identified. }
    \label{fig:injecting_l2_at_various_frequencies}
\end{figure}

Figure \ref{fig:injecting_l2_at_various_frequencies} shows the amplitude envelopes of several modes undergoing the emergent instability, with injection frequencies $\omega M_{i}$ at 0.4 (\textbf{a}) and $0.6$ (\textbf{b}). We first observe that the $2\omega$ component in $\ell=6$ undergoes a laminar excitation during the initial injection phase ($t<t_0=600M_{i}$, in gray). As discussed earlier in Sec.~\ref{subsec:parametric_instability_weak_regime}, it is a fourth-order effect: $2\omega(=\omega+\omega+\omega-\omega)$. The peak amplitude of this mode has been confirmed to scale as $A_{\rm inj}^4$. After the injection stabilizes $(t>t_0=600M_{i})$, $2\omega$ decays to some extent and undergoes the secondary excitation (open circle). The mode grows more rapidly for $\omega M_{i}=0.6$ than 0.4, and in the former case, the emergent excitation dominates over the initial laminar response. Ultimately, the evolution curve of  $2\omega$ forms a two-peak structure.  

The $\omega$ component in $\ell=6$ and $4$ follows a similar evolution pattern: it first undergoes a laminar rise (in gray), originating from a third-order coupling $(\omega=\omega+\omega-\omega)$, and is subsequently followed by a secondary excitation (open circle). In particular, the excitation of $\omega$ in $\ell=6$ occurs at much later times than $2\omega$ in the same harmonic.
These features are consistent with our earlier findings for $\omega M_{i}=0.5$.

The emergent instability, however, is strongly suppressed at higher frequencies. Figure~\ref{fig:injecting_l2_at_various_frequencies}\textbf{\textcolor{linkcolor}{c}} illustrates responses to an injection with $\omega M_{i} = 1.5$ and $A_{i}=10^{-3}$. As shown, all three modes exhibit an initial laminar rise but subsequently level off without any signs of further nonlinear growth. A likely explanation is that, in the high-frequency regime, the gravitational-wave driver becomes effectively decoupled from the black hole\footnote{Instead, the injected gravitational wave is directly absorbed by the black hole.}, making it more difficult to trigger the instability.

\begin{figure*}[!htb]
        \includegraphics[width=\linewidth]{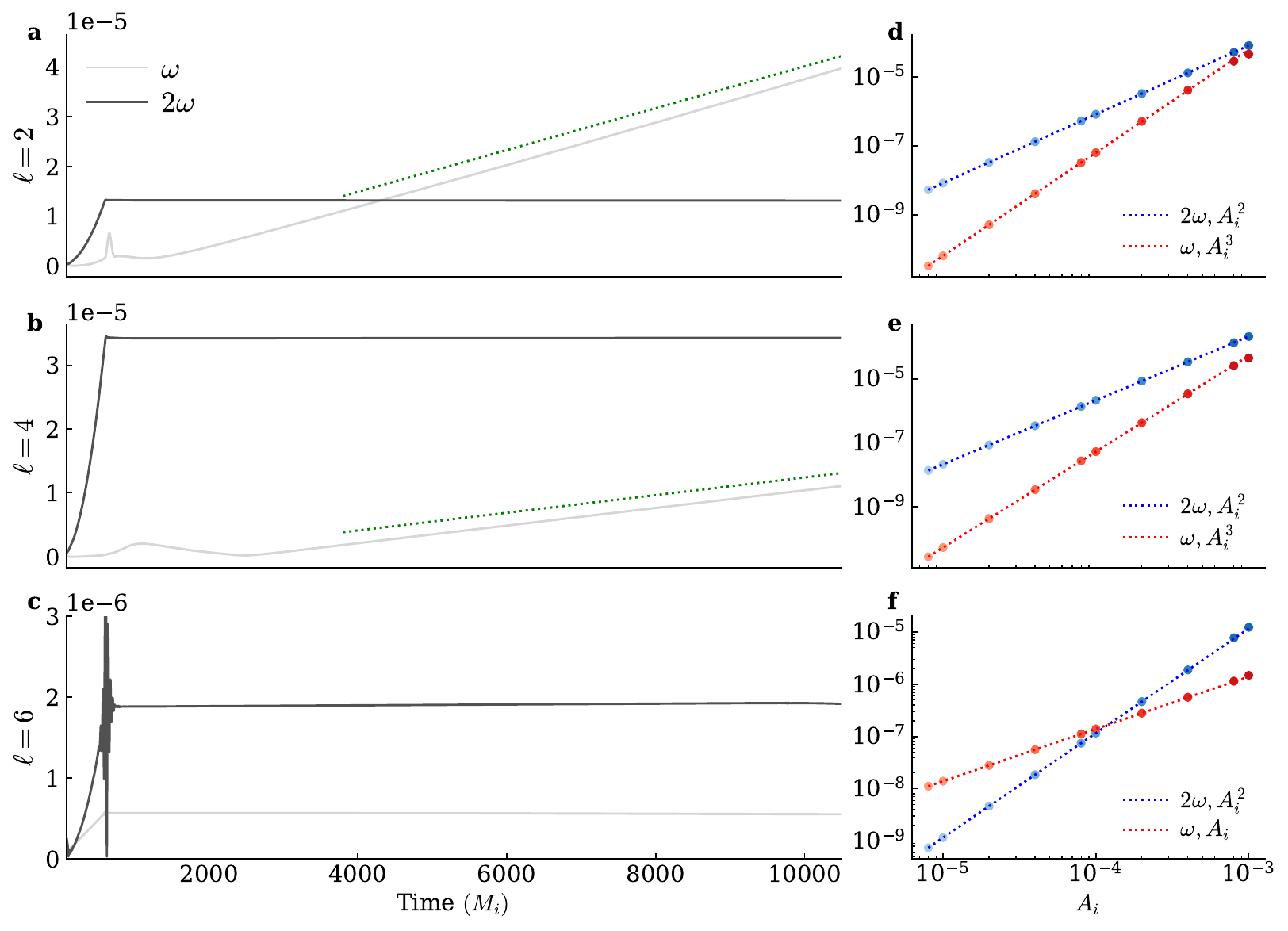}
    \caption{\textbf{Black hole's responses to a strong $\ell=6$ driver}. Responses are decomposed into two major constituent frequency components $\omega$ and $2\omega$, whose amplitude envelopes are shown. \textbf{a-c.} Responses in $\ell=2,4$, and 6 of $\mathcal{E}$, driven by a wave with frequency $\omega M_{i}=0.5$ and amplitude $A_{i}=4\times10^{-4}$. Green dotted lines indicate linear growth for reference. \textbf{d-f.} Measured growth rates or amplitudes of each mode as functions of injection amplitude $A_i$ in $\ell=2$ (\textbf{d}), $\ell=4$ (\textbf{e}), and $\ell=6$ (\textbf{f}). Dotted lines denote the $A_i^3,A_i^2$, and $A_i$ scaling, respectively, for comparison. }
    \label{fig:injecting_l6_response_4m4}
\end{figure*}

\section{Case study: $\ell=6$ driver}
\label{sec:}
We have shown that a three-mode coupling emerges from $\ell=6$ gravitational-wave drivers, leading to resonant excitations of $\omega$ in $\ell=2$ and 4 harmonics of the tendicity $\mathcal{E}$. This section further examines this scenario. 

\subsection{Simulations}
We first consider the Schwarzschild black hole case. Two dominant laminar modes include
\begin{itemize}
    \item Linear order: $\omega$ in $\ell=6$, with amplitude $\propto A_{i}$.
    \item Second order: $2\omega(=\omega+\omega)$ in $\ell=2,4,6$, with amplitude $\propto A_{i}^2$.
\end{itemize}
Figure \ref{fig:injecting_l6_response_4m4}\textbf{\textcolor{linkcolor}{a-c}} displays their time evolution, with injection amplitude $A_{i}=4\times10^{-4}$ and frequency $\omega M_{i}=0.5$. As expected, these modes react instantaneously to the injected wave, and their time evolution closely follows the behavior of the gravitational-wave driver. Once the driver settles, the mode amplitudes also stabilize at constant values, scaling as $A_{i}$ and $A_{i}^2$, respectively, as checked in Fig.~\ref{fig:injecting_l6_response_4m4}\textbf{\textcolor{linkcolor}{d-f}}. Consequently, their Fourier spectrum, shown in the main text, remains a fixed shape over time.

Notably, the linear mode $\omega$ in $\ell=6$ is 5 times weaker than the quadratic mode $2\omega$ in the same harmonic; and is $20-50$ times weaker than the quadratic modes $2\omega$ in $\ell=2$ and $\ell=4$. This feature contradicts expectations from standard perturbation theory, where linear responses typically dominate. This behavior can again be understood via the transmissivity curves in Fig.~\ref{fig:new_mass_change_suppl}\textbf{\textcolor{linkcolor}{b}}. 
In the present case $\omega M_{i}=0.5$, the transmissivity $|T_{\ell=6}|$ for the linear wave is extremely low, yielding a weak linear response in $\mathcal{E}$ $(< 10^{-7})$. By contrast, the quadratic modes $2\omega$ are governed by black hole second-order perturbation theory \cite{Campanelli:1998jv}, where the left-hand side of the master equation has the same form as the linear one, but the right-hand side includes a quadratic source at $2\omega$. As a result, the relevant transmissivity is still given by the linear result in Fig.~\ref{fig:new_mass_change_suppl}\textbf{\textcolor{linkcolor}{b}}, but should be evaluated at the sourcing frequency $2\omega$ instead. In $\ell=6$, the transmissivity of the quadratic source $2\omega$ is $2\times10^{4}$ times larger than that of the linear source, which more than compensates for the usual second-order suppression and makes the $2\omega$ mode stronger than the $\omega$ mode. Similarly, in $\ell=2,4$, the quadratic frequency lies above the corresponding quasinormal-mode frequencies (0.374 and 0.809). The quadratic modes can therefore fully tunnel through the $\ell=2$ and $\ell=4$ curvature potentials, leaving a stronger imprint in these harmonics.

Although $\omega$ barely penetrates the $\ell=6$ barrier, it nevertheless reaches the horizon through the third-order emergent instability in $\ell=2$ and 4. As discussed in the main text, the mode is resonantly excited via the three-mode coupling and exhibits linear growth in amplitude. This growth occurs across a range of injection amplitudes, with no evidence of a minimum threshold. As shown in Fig.~\ref{fig:injecting_l6_response_4m4}\textbf{\textcolor{linkcolor}{a-b}} (in gray), the $\ell=2$ mode grows rapidly and eventually dominates all other components at late times, illustrating an inverse energy cascade. In Fig.~\ref{fig:injecting_l6_response_4m4}\textbf{\textcolor{linkcolor}{d-e}}, we verify that the growth rate scales as $A_i^3$.

\begin{figure}[!h]
        \includegraphics[width=\linewidth]{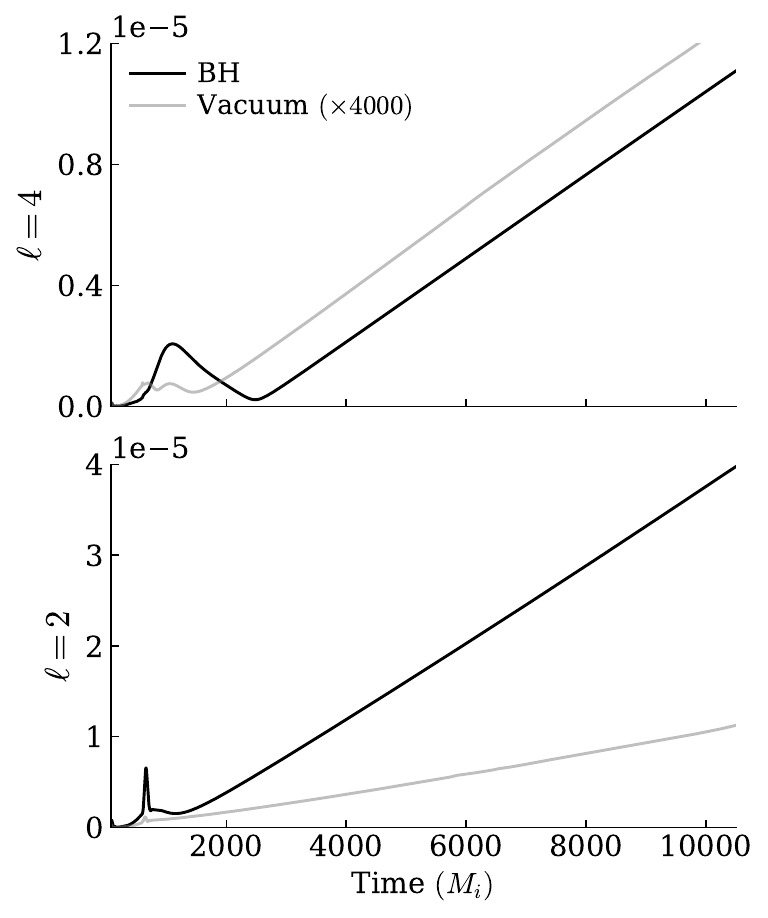}
    \caption{ Emergent instability of $\omega$ in $\ell=4$ (top panel) and $\ell=2$ (bottom panel) induced by an $\ell=6$ gravitational wave driver. Results are shown for both a Schwarzschild black hole (black) and Minkowski spacetime (gray). In the black hole case, responses are extracted on the apparent horizon; in the Minkowski case, they are measured on a coordinate sphere at radius 2. The Minkowski responses are magnified by a factor of 4000 to facilitate visual comparison. }
    \label{fig:vacuum_l2_l4_normalized_amp}
\end{figure}

According to the main text, the resonant excitation of $\omega$ in $\ell=2$ and 4 reflects a generic mode coupling in General Relativity, rather than a feature unique to black holes. To see this, we inject an $\ell=6$ gravitational wave into Minkowski spacetime, and measure $\mathcal{E}$ on a coordinate sphere at radius 2. The gray curves in Fig.~\ref{fig:vacuum_l2_l4_normalized_amp} show the amplitude envelopes of $\omega$ in $\ell=2$ and 4. Indeed, both exhibit linear growth in time. For comparison, we also provide the results for the black hole case (in black) with a comparable linear response --- that is, the amplitudes of $\omega$ in $\ell=6$ match between the black hole and vacuum injections. We can see that although the nonlinear resonance is present in both cases, the response on the black hole's horizon is three to four orders of magnitude stronger, demonstrating that the black hole substantially amplifies the nonlinear interaction.

These findings imply an interpretation that the observed resonance arises from the self-interaction of the injected gravitational wave, which gives rise to a persistent spacetime structure analogous to the concept of \emph{Geons}~\cite{PhysRev.97.511}. In the black hole scenario, the linear wave cannot tunnel through the curvature barrier and be absorbed; the spacetime structure is thus not destroyed by the black hole, allowing it to persist and drive the resonance.

\subsection{A toy model}
As discussed in the main text, the resonant behavior arises from a three-mode coupling. To illustrate this mechanism, consider a toy model in which a wave $\psi$ propagates on flat spacetime and is sourced by two parent modes $\psi_1$ and $\psi_2$. Then Eq.~(5) in the main text becomes
\begin{align}
    (\partial_t^2-\partial_x^2)\psi =\psi_1\psi_2. \label{eq:SM_toy_model}
\end{align}
Assume the parent modes take the form 
\begin{align}
    \psi_1=A_i^2 e^{2i\omega (t+x)}, \quad \psi_2=A_i e^{-i\omega (t+x)},
\end{align}
where $\psi_1$ represents a quadratic mode generated by the external driver. Eq.~\eqref{eq:SM_toy_model} then becomes 
\begin{align}
    (\partial_t^2-\partial_x^2)\psi =A_i^3 e^{i\omega (t+x)}. 
\end{align}
Performing a Fourier transform in the $x$-direction:
\begin{align}
    \psi(t,x)=\int \tilde{\psi}(t,k) e^{ikx} dk,
\end{align}
yields
\begin{align}
    \left(\frac{d^2}{dt^2}+k^2\right)\tilde{\psi}(t,k) =A_i^3 e^{i\omega t}\delta(k-\omega). 
\end{align}
This equation admits a resonant solution
\begin{align}
    \tilde{\psi}(t,k) = \frac{A_i^3 te^{i\omega t}}{2i\omega}\delta(k-\omega).
\end{align}
which, upon transforming back to real space, gives
\begin{align}
    \psi(t,x) = \frac{A_i^3 te^{i\omega (t+x)}}{2i\omega},
\end{align}
The amplitude of $\psi$ therefore grows linearly in time, demonstrating the resonant nature of the three-mode coupling.

\section{Simulation Methods}
We adopt the \texttt{SpEC} code \cite{SpECwebsite} to numerically evolve Einstein’s equations within a bounded computational domain.
The evolution is carried out within the Generalized Harmonic formalism \cite{Pretorius:2004jg,Lindblom:2005qh}, which casts the system into a set of symmetric hyperbolic partial differential equations. This formulation yields multiple characteristic modes, and boundary conditions are imposed only on those propagating into the computational domain. At the outer boundary, the required boundary conditions can be categorized into three subsets: constraint, gauge, and physical. In the discussion below, we will focus on the physical part, which encodes incoming gravitational radiation.

In \texttt{SpEC}, the Bjørhus method is used to impose boundary conditions. The physical subset reads \cite{Kidder:2004rw,Lindblom:2005qh,Buchman:2024zsb}
\begin{align}
    P_{\mu\nu}^{{\rm P}\alpha\beta} d_t u^{1-}_{\alpha\beta} \doteq \left({\rm incoming~ gravitational~wave }\right)_{\mu\nu}, \label{eq:physical_bc_1}
\end{align}
where the characteristic field $u^{1-}_{\mu\nu}$ is given by
\begin{align}
u^{1-}_{\mu\nu}=\Pi_{\mu\nu}-s^i\Phi_{i\mu\nu}-\gamma_2\psi_{\mu\nu}.
\end{align}
Here $s^i$ is the outward unit normal vector of the boundary. The projection operator $P_{\mu\nu}^{{\rm P}\alpha\beta}$, defined as \cite{Kidder:2004rw,Lindblom:2005qh}
\begin{equation}
\begin{split}
    &P_{\mu\nu}=g_{\mu\nu}+n_{\mu}n_{\nu}-s_{\mu}s_{\nu}, \\
    &P_{\mu\nu}^{{\rm P}\alpha\beta}=P_{\mu}{}^{\alpha}P_{\nu}{}^{\beta}-\frac{1}{2}P_{\mu\nu}P^{\alpha\beta},
\end{split}
\end{equation}
singles out the two physical degrees of freedom associated with gravitational waves. For our simulations, we adopt the default choice of the constraint-damping parameter $\gamma_2$ commonly used in binary black hole evolutions.

To implement the boundary conditions in Eq.~\eqref{eq:physical_bc_1}, a convenient approach is to set the value of $P_{\mu\nu}^{{\rm P}\alpha\beta} d_t u^{1-}_{\alpha\beta}$ at the boundary via \cite{Lindblom:2005qh}
\begin{align}
    &P_{\mu\nu}^{{\rm P}\alpha\beta} d_t u^{1-}_{\alpha\beta} \doteq P_{\mu\nu}^{{\rm P}\alpha\beta} \notag \\
    &\times \left[D_tu^{1-}_{\alpha\beta}-(\alpha+s_i\beta^i)(w_{\alpha\beta}^{-}-\left. w_{\alpha\beta}^{-}\right|_{\rm BC}-\gamma_2s^ic_{i\alpha\beta}^{\hat{3}})\right],
\end{align}
where $\alpha$ and $\beta^i$ are the lapse and shift, respectively. The quantities $D_tu^{1-}_{\alpha\beta}$ and $c_{i\alpha\beta}^{\hat{3}}$ are not relevant for our purposes; their definitions can be found near Eqs.~(57) and (62) of \cite{Lindblom:2005qh}.
The tensor $w_{\alpha\beta}^{-}$ is constructed from the Weyl tensor $C_{\mu\rho\nu\tau}$ as
\begin{align}
    w_{\alpha\beta}^{-}= P_{\alpha\beta}^{{\rm P}\mu\nu}(n^\rho-s^\rho)(n^\tau-s^\tau)C_{\mu\rho\nu\tau}.
\end{align}
The term $\left. w_{\alpha\beta}^{-}\right|_{\rm BC}$ denotes the target value of $w_{\alpha\beta}^{-}$ at the boundary and serves as the physical input controlling incoming gravitational waves throughout a simulation. 

To inject gravitational waves into a system, one simply prescribes a desired $\left. w_{\alpha\beta}^{-}\right|_{\rm BC}$ as a function of time and angular coordinates. Importantly, this prescription does not violate any of the constraints. This feature allows one to continuously inject gravitational waves --- a task that is challenging to achieve using initial data alone, where the Hamiltonian and momentum constraint equations must be solved in the presence of (effectively) infinite-duration gravitational waves.

Although $\left. w_{\alpha\beta}^{-}\right|_{\rm BC}$ is a symmetric spacetime tensor, it carries only two dynamical degrees of freedom. This can be seen explicitly by writing \cite{Ma:2023qjn}
\begin{align}
    \left. w_{\alpha\beta}^-\right|_{\rm BC}=2(\Psi_0 \bar{m}_\alpha \bar{m}_\beta+\bar{\Psi}_0m_\alpha m_\beta),
\end{align}
where the covariant vector $\bm{m}$ reads
\begin{align}
    \bm{m}  = \frac{R_{\rm Bdry}}{\sqrt{2}} (d\theta+i\sin\theta d\phi),
\end{align}
and $R_{\rm Bdry}$ is the coordinate radius of the outer boundary. Thus, all dynamical information about the incoming radiation is encoded in the complex Weyl scalar $\Psi_0$, which represents the two polarization modes of gravitational waves. This representation forms the core foundation for implementing Cauchy-characteristic matching \cite{Ma:2023qjn,Ma:2024hzq}, where $\Psi_0$ is computed from a separate characteristic system that covers the wave zone. In our context, as discussed earlier, we 
perform a modal decomposition of $\Psi_0$ as
\begin{align}
    \Psi_0(t,\theta,\phi)=\sum_{\ell,m}A_{\ell m}(t) \times {}_{+2}Y_{\ell m}(\theta,\phi).
\end{align}
The real (imaginary) part of $A_{\ell m}(t)$ corresponds to a gravitoelectric (gravitomagnetic) tidal field. To leading order in a Schwarzschild black hole, these represent polar (even-parity) and axial (odd-parity) 
perturbations, respectively \cite{chandrasekhar1998mathematical}.
Throughout this study, we consider non-rotating black holes and drive them exclusively with purely electric (even-parity) waves.

During simulations, the inner excision boundary is dynamically adjusted using the adaptive control system \cite{Hemberger:2012jz}, based on the dual-frame method \cite{Scheel:2006gg}. Truncation errors are managed via the adaptive mesh refinement algorithm \cite{Szilagyi:2014fna}.

\end{document}